\documentclass[aps,prx,twocolumn,superscriptaddress,nofootinbib,a4paper,longbibliography]{revtex4-2}
\pdfoutput=1
\usepackage[utf8]{inputenc}
\usepackage[T1]{fontenc}
\usepackage[english]{babel}
\usepackage{amsmath,amssymb,amsthm}
\usepackage{microtype}
\usepackage{amsfonts}
\usepackage{amsmath}
\usepackage{amsthm}
\usepackage{amssymb}					
\usepackage{amsthm}
\usepackage{dsfont}
\usepackage{bm}
\usepackage{mathtools}
\usepackage{color}
\usepackage{multirow}
\usepackage[normalem]{ulem}
\usepackage{latexsym}
\usepackage{mathrsfs}
\usepackage{natbib}
\usepackage{verbatim}
\usepackage[T1]{fontenc}
\usepackage{float}
\usepackage{graphicx}
\usepackage{xcolor}
\usepackage{physics}
\usepackage{subcaption}

\usepackage{caption}
\newcommand{\zs}{{\mathtt{S}}}
\newcommand{\zl}{{\mathtt{L}}}

\usepackage[colorlinks=true,linkcolor=blue,citecolor=magenta,urlcolor=blue]{hyperref}

\newcommand{\A}{{\mathsf{A}}}
\newcommand{\B}{{\mathsf{B}}}
\newcommand{\T}{{\mathsf{T}}}

\begin{document}


\title{Device-Independent Quantum Key Distribution Based on Routed Bell Tests}

\author{Tristan Le Roy-Deloison}
\affiliation{Univ Lyon, Inria, ENS Lyon, UCBL, LIP, Lyon, France}
\affiliation{Télécom Paris-LTCI, Institut Polytechnique de Paris, Palaiseau, France}

\author{Edwin Peter Lobo}
\affiliation{Laboratoire d'Information Quantique, Universit\'e libre de Bruxelles (ULB), Belgium}

\author{Jef Pauwels}
\affiliation{Department of Applied Physics, University of Geneva, 1211 Geneva, Switzerland}
\affiliation{Constructor Institute of Technology (CIT), 1211 Geneva, Switzerland}
\affiliation{Constructor University, 28759 Bremen, Germany}

\author{Stefano Pironio}
\affiliation{Laboratoire d'Information Quantique, Universit\'e libre de Bruxelles (ULB), Belgium}

\begin{abstract}
    Photon losses are the main obstacle to fully photonic implementations of device-independent quantum key distribution (DIQKD). Motivated by recent work showing that routed Bell scenarios offer increased robustness to detection inefficiencies for the certification of long-range quantum correlations, we investigate DIQKD protocols based on a routed setup. In these protocols, in some of the test rounds, photons from the source are routed by an actively controlled switch to a nearby test device instead of the distant one. We show how to analyze the security of these protocols and compute lower bounds on the key rates using noncommutative polynomial optimization and the Brown-Fawzi-Fawzi method. We determine lower bounds on the asymptotic key rates of several simple two-qubit routed DIQKD protocols based on CHSH or BB84 correlations and compare their performance to standard protocols. For high-quality short-path tests, we find that routed DIQKD protocols are significantly more robust to losses, showing an improvement of approximately $30\%$ in the detection efficiency compared to their nonrouted counterparts. This translates to a large improvement in the distance over which nonzero key can be distilled in optical setups with near-perfect single-photon detectors, where the main source of loss in the setup is due to transmission in the fiber. Notably, the routed BB84 protocol achieves a positive key rate with a detection efficiency as low as $50\%$ for the distant device, the minimal threshold for any QKD protocol featuring two untrusted measurements. 
    However, the advantages we find are highly sensitive to noise and losses affecting the short-range correlations involving the additional test device. 
\end{abstract}


\maketitle

\section{Introduction}
A prerequisite for device-independent quantum information protocols is the ability to certify genuine quantum correlations between the devices involved \cite{Acin2007,Brunner2014}. In practice, this requires the ability to perform Bell tests free of the detection loophole \cite{Pearle1970,Clauser1974,Garg1987}, i.e., to detect quantum particles with a high enough efficiency. In full photonic implementations, this is one of the main obstacles to overcome because of unavoidable losses in the quantum optical channel. In particular, the distance record for full photonic loophole-free Bell tests is of the order of 200 m \cite{Shalm2021,Li2021,Liu2021}. This is far from the distances required for practical applications. 

One possible way to overcome optical losses and reach high enough detection efficiencies is to use an `event-ready' scheme \cite{Bell2004}, where the presence of entanglement between the two remote devices is heralded before they perform their measurements. This may be achieved through entanglement swapping \cite{Zukowski1993,Sangouard2011,Curty2011}, quantum amplifier \cite{Gisin2010}, local precertification of the photons \cite{Cabello2012}, or full quantum repeaters \cite{Azuma2023}. In each case, this requires additional sources of quantum particles and/or joint measurements, substantially increasing the implementation complexity.

\begin{figure}[t]
    \centering
    \begin{subfigure}{0.485\textwidth}
    \centering
    \includegraphics[width=8.6cm]{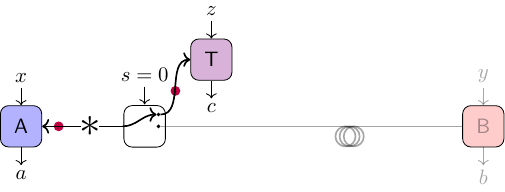}
     \caption{Bob's particle is routed to the closeby device when $s=0$.}
    \end{subfigure}
    \hfill
    \begin{subfigure}{0.485\textwidth}
    \centering
    \includegraphics[width=8.6cm]{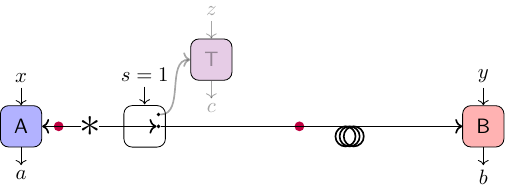}
    \caption{Bob's particle is routed to the distant device when $s=1$.}
    \end{subfigure}
    \caption{\label{fig:setup} The routed Bell scenario. }
\end{figure}

Recently, the idea of routed Bell tests has been proposed \cite{Chaturvedi2022,Lobo2023} as a simple modification to standard Bell tests that can reduce the detection efficiency required for loophole-free experiments, and hence extend the distance over which quantum correlations can be certified. The basic idea behind a routed Bell test is depicted in Fig.~\ref{fig:setup}. As in a standard Bell test, it features a measurement device $\mathsf{A}$ for Alice, a measurement device $\mathsf{B}$ for Bob, and a source of entangled particles. In each experimental trial, Alice and Bob can operate their devices independently, submitting random inputs $x$ and $y$, respectively, (the measurement settings), and obtaining classical outputs $a$ and $b$. However, in the routed configuration, an additional element is introduced: the possibility for Bob's particle to be routed via a switch to a different measurement device than $\mathsf{B}$. We denote this additional device $\mathsf{T}$ and its corresponding input and output $z$ and $c$. The switch is controlled by a classical input $s$, which determines whether Bob's particle is routed to $\mathsf{T}$ or $\mathsf{B}$. For example, $s=0$ routes it to $\mathsf{T}$ and $s=1$ routes it to $\mathsf{B}$.  Depending on the switch setting $s$, one can thus either perform a Bell test in the $\A/\T$ configuration or the $\A/\B$ configuration. The purpose of the $\A/\T$ test is to certify, as best as possible, the quantum behavior of the particle source and of the device $\mathsf{A}$. To minimize losses in these tests, and reach a high Bell violation, the devices $\mathsf{A}$ and $\mathsf{T}$ are thus situated close to the source of entangled particles. Performing a Bell test with the $\A/\T$ setup for a randomly selected subset of the trials will then ensure that Alice's device $\mathsf{A}$ behaves (almost) honestly, even when it is part of the long-distance $\A/\B$ test. This limits how $\mathsf{A}$ can collude with $\mathsf{B}$ to simulate genuine quantum correlations, thereby lowering the detection efficiency threshold required to authenticate such correlations in the $\A/\B$ configuration.

As in standard DIQKD, all components of the setup, including the switch and the additional measurement device $\mathsf{T}$ are untrusted and their internal functioning is uncharacterized. The only assumption made on the devices is that they obey certain no-signaling constraints preventing them from signaling arbitrarily to each other. Specifically, the behavior of the devices on Alice's side of the entangled source should not influence the devices on Bob's side, and vice versa. Thus, in a given trial, the classical input $x$ and output $a$ on Alice's side should not influence Bob's particle and the measurements performed at $\mathsf{T}$ or $\mathsf{B}$. Similarly, the classical inputs and outputs $s, z, c, y,$ and $b$ on Bob's side should not influence Alice's quantum particle or the measurement performed at $\mathsf{A}$. This last condition is crucial to ensure that the source and the measurement device $\mathsf{A}$ behave identically whether Bob's particle is routed to $\mathsf{B}$ or $\mathsf{T}$. This is necessary to ensure that the $\A/\T$ tests are a reliable indicator of the behavior of $\mathsf{A}$ also in the $\A/\B$ configuration.

In Ref.~\cite{Lobo2023}, criteria and tools are introduced to certify long-range quantum correlations between $\mathsf{A}$ and $\mathsf{B}$, given that certain correlations are observed in the short-range $\A/\T$ test. Although the required detection efficiency at $\mathsf{B}$ can be lowered compared to standard Bell tests, which lack the intermediate $\A/\T$ test, the improvements  are highly sensitive to losses in the $\A/\T$ test. 

Consequently, routed Bell tests provide a promising avenue for extending the distance over which long-range quantum correlations can be certified if high-quality devices are used in the implementation. To illustrate this point, let us write the detection efficiency as $\eta = \eta_f \eta_d$, where $1-\eta_d$ denotes the losses from transmission, which increases exponentially with distance, whereas $1-\eta_f$ is a fixed value that captures all other sources of loss, such as the local detectors, the switch, and the source of entangled photons. For the devices $\A$ and $\T$ close to the source, $1-\eta_d$ is negligible. When $1-\eta_f$ is negligible for all devices $\A, \B$ and $\T$, losses are dominated by $\eta_d$, and routed Bell tests can demonstrate long-range quantum correlations for a lower value of $\eta_d$ for $\B$ than standard Bell tests, thereby extending the distance over which quantum correlations can be certified. While experiments with negligible $1-\eta_f$ are still out of reach for current technology, this is expected to significantly improve with future progress. In contrast, the exponential decrease of $\eta_d$ with distance is a fundamental limitation that arises due to the nature of interaction between the photons and the fiber.

Further, the routed Bell scenario exhibits features that could be of interest to DIQKD. For instance, the BB84 correlations, which are emblematic in QKD, and can be produced from a maximally two-qubit state $|\phi_+\rangle$ by carrying out $\sigma_z$ and $\sigma_x$ Pauli measurements at $\mathsf{A}$ and $\mathsf{B}$, can be replicated classically in a standard Bell setup, hence are unsuitable for standard DIQKD. However, their quantum nature can be certified in a routed Bell experiment by performing random CHSH tests in the $\A/\T$ configuration, with the testing device $\mathsf{T}$ using the CHSH bases $(\sigma_z\pm \sigma_x)/\sqrt{2}$ \cite{Lobo2023}. Furthermore, when the short-distance $\A/\T$ test achieves the maximal CHSH value, the detection efficiency threshold for the device $\mathsf{B}$ is $50\%$, the minimal one for a two-measurement device \cite{Massar2003}.

Alice's measurements at $\mathsf{A}$ can also be seen as remotely preparing states for Bob's device $\B$. The $\A/\T$ tests within a routed Bell framework self-test the entangled particle source and Alice's device $\mathsf{A}$, hence they self-test those remotely prepared states.  When this self-test is perfect, these remotely prepared states are fully characterized. Hence, the additional measurements made at $\mathsf{T}$ effectively turn a DIQKD protocol into a one-sided DI prepare-and-measure QKD protocol, which is typically more efficient and noise robust. 

The above observations, and the relative simplicity of implementing routed Bell tests, motivate the study of their applications to DIQKD, both for improving the experimental prospects for DIQKD and our conceptual interest.
In the present paper, we introduce simple DIQKD protocols based on routed Bell tests and show how lower bounds on the key rates can be computed numerically using noncommutative polynomial optimization (NPO) \cite{Navascues2007,Navascues2008,Pironio2010a} and the BFF method \cite{Brown2021}. We determine such lower bounds for several simple routed DIQKD (rDIQKD) protocols and compare their performance to standard DIQKD protocols.

\section{Routed DIQKD}
\subsection{Setup}
The rDIQKD protocols that we introduce are based on the routed Bell configuration, represented in Fig.~\ref{fig:setup} and succinctly described in the Introduction. As in standard DIQKD, we assume that Alice and Bob are in safe laboratories and that they each operate independently the untrusted measurement devices $\mathsf{A}$ and $\mathsf{B}$, respectively. In most of the rounds, the outputs of these devices are kept secret and will constitute the raw key. In the remaining rounds, the outputs are publicly announced and contribute to statistical data collection for parameter estimation.

Part of the time, Bob's particle is also randomly routed to the testing device $\mathsf{T}$ instead of $\mathsf{B}$. The outputs of such rounds are never used to generate a key but always contribute to parameter estimation. As regards the location of the device $\mathsf{T}$, as well as of the entangled source and the switch, there are then two possibilities.

\begin{figure}[t]
    \centering
    \begin{subfigure}{0.485\textwidth}
        \centering
        \includegraphics[width=8.6cm]{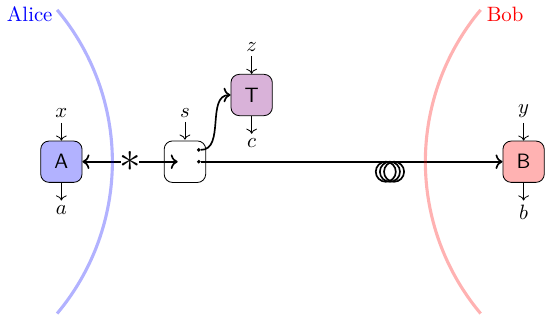}
         \caption{rDIQKD setup assuming the switch and the device $\mathsf{T}$ are outside Alice's and Bob's safe laboratories.\label{fig:rdiqkd-setup-a}}
        \end{subfigure}
        \hfill
        \begin{subfigure}{0.485\textwidth}
        \centering
        \includegraphics[width=8.6cm]{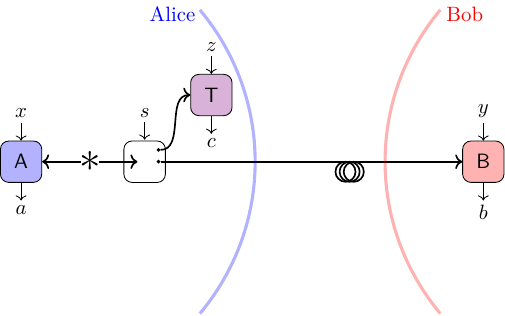}
        \caption{rDIQKD setup where the switch and the device $\mathsf{T}$ are located in Alice's laboratory.\label{fig:diqkd-setup-b}}
        \end{subfigure}
    \caption{\label{fig:rdiqkd-setup} Two possible setups for rDIQKD.}
\end{figure}

The first, depicted in Fig.~\ref{fig:rdiqkd-setup-a}, is to consider them as being outside Alice's and Bob's laboratories and in full control of the eavesdropper, similar to the entangled source in standard entanglement-based QKD or the joint measurement in measurement-device-independent QKD. We can then think\footnote{The alternative possibility of incorporating a trusted intermediary between $A$ and $B$ to supply random inputs to the switch and device $\mathsf{T}$ departs from our focus, which is on direct, secure communication between Alice and Bob. Furthermore extending QKD at a large distance through third-party trusted intermediaries is always possible, but opens new additional points of potential vulnerability.} of, say, Alice as providing through public announcements the classical inputs $s$ and $z$, and the device $\mathsf{T}$ as answering back with a publicly announced classical output $c$. The difficulty with this setup is that we should ensure that, upon learning $s$ or $z$, the eavesdropper does not modify Alice's quantum state before it enters her laboratory and her measurement is completed, as this would violate the no-signaling requirements discussed in the Introduction. In principle, Alice could announce the value of $s$ or $z$ after her measurement is completed and she has recorded her output $a$. However, this would require delaying the operation of the switch until that moment, which in practice would require the use of a quantum memory to store the quantum state until the announcement is made. This is a significant experimental complication.

The other option is to assume that the switch and the device $\mathsf{T}$ are all situated in Alice's laboratory, as depicted in Fig.~\ref{fig:diqkd-setup-b}, and that these devices cannot arbitrarily communicate their private inputs to Alice's device $\mathsf{A}$. This is a customary assumption in DI quantum cryptography \cite{Pironio2010}, which can, e.g., be enforced through shielding of the devices. This is a setup that amounts effectively to viewing Alice as holding a measurement-based preparation device that prepares quantum states that are sent to Bob on an untrusted public quantum channel; part of the time, these states are not sent by Alice on the public channel but are instead locally measured by $\mathsf{T}$ in her laboratory to verify their quantum properties. We stress that even though in this setup we assume that we can control and restrict the classical and quantum communication that goes in and out of the devices involved in the protocol, they are all still viewed as black boxes whose internal functioning is untrusted.

Although it is more natural to think of rDIQKD as implemented in the setup of Fig.~\ref{fig:diqkd-setup-b}, we will analyze in the following its security in the setup of Fig.~\ref{fig:rdiqkd-setup-a}, where Eve can freely control and access the internal state of the switch and the device $\mathsf{T}$. Clearly, security in this later case also implies security in the case of Fig.~\ref{fig:diqkd-setup-b}, as we are giving more power to Eve.

\subsection{Generic protocols}
We now outline the general procedure of a generic rDIQKD protocol based on the setup described above. In each measurement round $i=1,\ldots,n$ of the protocol, Alice generates independent random variables $X_i\in\mathcal{X}$ and $S_i\in\{0,1\}$ and feeds them, respectively, to her device $\mathsf{A}$ and the switch. If $S_i=0$, indicating the choice to route Bob's quantum particle to $\mathsf{T}$, she also generates a random variable $Z_i\in\mathcal{Z}$ and feeds it to $\mathsf{T}$. She records the output variables $A_i\in\mathcal{A}$ and, if applicable, $C_i\in\mathcal{C}$. She publicly announces $S_i$. If $S_i=1$, indicating the choice to route Bob's quantum particle to $\mathsf{B}$, Bob generates a random variable $Y_i\in\mathcal{Y}$, feeds it to his device $\mathsf{B}$, and records the output $B_i\in\mathcal{B}$. Alice and Bob then start a new round $i\rightarrow i+1$.

Once all $n$ measurement rounds have been completed, Alice and Bob communicate on a public classical channel with two main goals. On one hand, they disclose part of the data they generated to check that a statistical test $\Gamma$ is passed, such as verifying a significant violation of a routed Bell inequality. On the other hand, they agree on a subset of the rounds for which they will keep the variables $A_i$ and $B_i$ secret. These variables will constitute the raw key. Typically, these key generation rounds will be those for which the inputs of Alice and Bob belong to a certain subset $\mathcal{K}$ of all their possible input pairs $\mathcal{X}\times\mathcal{Y}$ (e.g., they may generate a key only when Alice uses input $x=0$ and Bob uses input $y=3$). 

The probabilities with which the input variables are chosen in the $n$ measurement rounds are usually fixed to maximize the key rate, while at the same time ensuring that enough data is obtained for the test $\Gamma$ to be statistically significant. One might for instance choose these probabilities so that the number of key generation rounds is roughly of order $n-\sqrt{n}$, while the number of rounds used for parameter estimation is of order $\sqrt{n}$. 

Finally, if the statistical test $\Gamma$ is passed, Alice and Bob apply error correction and privacy amplification techniques to their copy of the raw key to extract the finally shared secret key.

\subsection{Long-range quantum correlations are necessary for security}
Before explaining how the security of rDIQKD protocols can be analyzed and key rates computed, we first point out that long-range quantum correlations, as defined in Ref.~\cite{Lobo2023}, are necessary for the security of rDIQKD protocols, in the same way that nonlocal correlations are necessary for the security of standard DIQKD protocols.

Routed Bell scenarios feature a short-range Bell test, involving the $\A/\T$ devices, and a long-distance Bell test, involving the $\A/\B$ devices. In any given round, the quantum strategy that is used gives rise to the correlations
\begin{equation}\label{eq:corr}
    \begin{cases}
        p(a,c|x,z) = \tr\left(\rho_{AB}\,\mathsf{A}_{a|x} \otimes \mathsf{T}_{c|z}\right) & \text{if } s=0\,,\\
        p(a,b|x,y) = \tr\left(\rho_{AB}\,\mathsf{A}_{a|x}\otimes \mathsf{B}_{b|y}\right) & \text{if } s=1\,,
        \end{cases}
\end{equation}
where $\rho_{AB}$ is the entangled state produced by the sources and $\mathsf{A}_x=\{\mathsf{A}_{a|x}\}_a$, $\mathsf{T}_z=\{\mathsf{T}_{c|z}\}_c$ and $\mathsf{B}_y=\{\mathsf{B}_{b|y}\}_b$ are the POVMs performed by the devices $\mathsf{A}$, $\mathsf{T}$ and $\mathsf{B}$, respectively. 

Following Ref.~\cite{Lobo2023}, we say that the correlations $\{p(a,c|x,z), p(a,b|x,y)\}$ are short-range quantum (SRQ) correlations if they can be simulated by a quantum model in which the measurements performed at $\mathsf{B}$ are jointly measurable. Formally, this means that the correlations can be written as 
\begin{equation}\label{eq:srq}
    \begin{cases}
        p(a,c|x,z) = \tr\left(\tilde\rho_{AB}\,\tilde{\mathsf{A}}_{a|x} \otimes \tilde{\mathsf{T}}_{c|z}\right) & \text{if } s=0\,,\\
        p(a,b|x,y) = \sum_\lambda p(b|y,\lambda)\,\tr\left(\tilde\rho_{AB}\,\tilde{\mathsf{A}}_{a|x}\otimes \mathsf{N}_{\lambda}\right) & \text{if } s=1\,,
    \end{cases}
\end{equation}
for some quantum state $\tilde\rho_{AB}$, POVMs $\tilde{\mathsf{A}}_{x}=\{\tilde{\mathsf{A}}_{a|x}\}_a$ for $\mathsf{A}$,  $\tilde{\mathsf{T}}_{z}=\{\tilde{\mathsf{T}}_{c|z}\}_c$ for $\mathsf{T}$, and a single `parent' POVM $\mathsf{N}=\{\mathsf{N}_\lambda\}_\lambda$ for $\mathsf{B}$, independent of the input $y$. The outcome $b$ at $\mathsf{B}$ is then outputted with probability $p(b|y,\lambda)$ depending on the classical outcome $\lambda$. Given that the parent POVM $\mathsf{N}$ is independent of $y$, it can be carried out near the switch, eliminating the need to transmit Bob's quantum particle to the distant device $\mathsf{B}$. Instead, only the classical outcome $\lambda$ of this parent measurement needs to be communicated to $\mathsf{B}$. Quantum correlations that cannot be written as in Eq.\eqref{eq:srq} are called long-range quantum (LRQ) correlations in Ref.~\cite{Lobo2023}. These require the transmission of quantum information to $\mathsf{B}$ for a genuine quantum measurement to occur after the input $y$ to $\mathsf{B}$ is provided. 

In our rDIQKD context, if the quantum devices of Alice and Bob only generate SRQ correlation of the form Eq.~\eqref{eq:srq}, then the parent POVM $\mathsf{N}$ can be performed by Eve on the public channel between the switch and the device $\mathsf{B}$. As Eve can keep a copy of the classical outcome $\lambda$, the correlations between $A$ and $B$ factorize when conditioned on Eve's information, i.e., 
\begin{equation*}
    \begin{aligned}
        p(a,b|x,y,\lambda) &=  \frac{\tr\left(\tilde\rho_{AB}\,\tilde{\mathsf{A}}_{a|x}\otimes \mathsf{N}_{\lambda}\right)}{\tr\left(\tilde\rho_{AB}\,\mathbb{I}\otimes \mathsf{N}_{\lambda}\right)} \, p(b|y,\lambda)\, ,\\
        &=p(a|x,\lambda) p(b|y,\lambda) \, ,
    \end{aligned}
\end{equation*}
where $p(a|x,\lambda) \linebreak[1]=  \tr\left(\tilde\rho_{AB}\,\tilde{\mathsf{A}}_{a|x}\otimes \mathsf{N}_{\lambda}\right)/ \tr\left(\tilde\rho_{AB}\,\mathbb{I}\otimes \mathsf{N}_{\lambda}\right)$, implying that no secure key can be extracted \cite{maurer1999unconditionally}.

In Ref.~\cite{Lobo2023}, techniques are introduced for determining when a given set of correlations is SRQ and minimal detection efficiencies required to exhibit LRQ correlations are presented for various cases. These results put constraints on the required detection efficiencies for rDIQKD. For instance, it is shown that in a routed Bell scenario, where Alice and Bob have two inputs, i.e., $\mathcal{X}=\mathcal{Y}=\{0,1\}$, no LRQ correlations can be generated if the detection efficiencies $\eta_\A$ of $\mathsf{A}$ and $\eta_\B$ of $\mathsf{B}$ satisfy
\begin{equation}\label{eq:bound-model}
    \eta_{\B}\leq \frac{\eta_\A}{3\eta_\A-1}\,.
\end{equation}
For instance, when $\eta_\A=1$, no key can be extracted if $\eta_\B\leq 1/2$.

\subsection{Proving security of rDIQKD protocols}

As in standard DIQKD, and other QKD protocols, proving the security of an rDIQKD protocol amounts to finding, given that the statistical test $\Gamma$ is passed, a lower bound on the smooth min-entropy $H_\text{min}^\epsilon(A^n|E)$ of Alice's final raw string $A^n=A_1\ldots A_n$ conditioned on the eavesdropper's information $E$, which includes all information publicly disclosed in the protocol as well as Eve's quantum side information acquired during the protocol by interacting with Alice's and Bob's quantum systems. Provided the bound is sufficiently high, it guarantees that privacy amplification can be performed to extract a secure key of the desired length.

The security of standard DIQKD protocols composed of $n$ measurement rounds, which may generally not follow an independent and identically distributed (IID) model and where memory effects can be present in the devices, can be reduced to a single-round analysis through the use of the entropy accumulation theorem (EAT) \cite{dupuis2020entropy,arnon2018practical,arnon2019simple} or the generalized entropy accumulation theorem (GEAT) \cite{Metger2022}. One then finds that the smooth min-entropy is basically the same, up to sublinear terms in $n$, as in the case where the devices behave identically and independently in each round of the protocol. That is, roughly, $H_\text{min}^\epsilon(A^n|E)\geq n H(A|E)-O(\sqrt{n})$ where $H(A|E)$ is the conditional von Neumann entropy of Alice's output $A$ given Eve's information $E$ in a single i.i.d. round. In particular, the asymptotic key rate (when $n\rightarrow\infty$) of the protocol against the most general attacks is the same as the IID key rate given by the Devetak-Winter bound \cite{Devetak2005}.

To apply the EAT or GEAT following the approach of Refs.~\cite{arnon2018practical,arnon2019simple}, the CPTP maps $\mathcal{P}_i$ describing the individual steps of an intermediary `entropy accumulation' (EA) protocol, to which the security of the entire protocol can be reduced, must satisfy a certain Markov condition, in the case of EAT, or a no-signaling condition, in the case of the GEAT. These conditions capture the idea that any side information Eve may hold about the measurement results at step $i$ is outputted at step $i$ and not recorded in Alice or Bob's quantum systems to be later passed out to Eve in a subsequent round. This is essentially the only nontrivial technical requirement that has to be satisfied for applying either the EAT or GEAT. 

An rDIQKD protocol mainly differs from a DIQKD protocol only in the purely quantum phase of the protocol, where Alice and Bob's entangled systems are measured to produce classical outputs $A_i$, $B_i$, and $C_i$. Other aspects of the protocol, like public communication, sifting, parameter estimation, error correction, etc, are essentially identical and can be analyzed in the same way. In particular, the reduction from a multiround analysis to a single-round analysis using the EAT or GEAT can be done following the approach outlined in Refs.~\cite{arnon2018practical,arnon2019simple} for DIQKD protocols. The sole requirement is to verify that the no-signaling condition necessary for applying the GEAT is satisfied. 

\begin{figure}[t]
    \centering
    \includegraphics[width = 8.6cm]{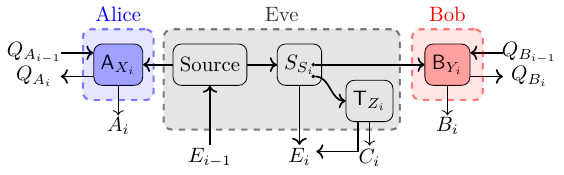}
    \caption{\label{fig:geat} Structure of the CPTP maps $\mathcal{M}_i$ describing the quantum measurement phase of the protocol at step $i$. }
\end{figure}

\begin{figure}[t]
    \centering
    \begin{subfigure}{0.485\textwidth}
        \centering
        \includegraphics[width=\textwidth]{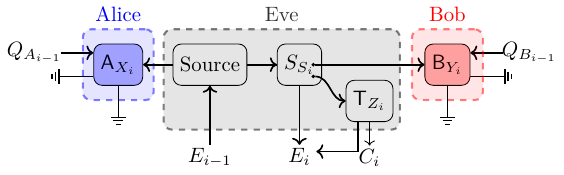}
        \caption{}
        \end{subfigure}
        \hfill
        \begin{subfigure}{0.485\textwidth}
        \centering
        \includegraphics[width=\textwidth]{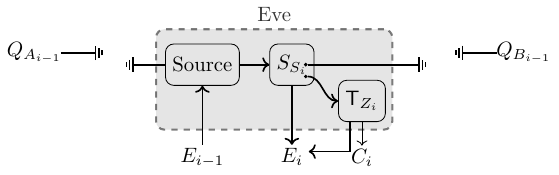}
        \caption{}
        \end{subfigure}
        \caption{\label{fig:geat-traced}(a) The map  $\tr_{A_iB_iQ_{A_i}Q_{B_i}}\circ \mathcal{M}_i $ obtained by tracing out the map in Fig.~\ref{fig:geat} over the output systems of Alice and Bob.  As depicted in (b), it does not depend anymore on the input registers $Q_{A_{i-1}}$ and $Q_{B_{i-1}}$, i.e., it is of the form $\mathcal{E}_i\circ \tr_{Q_{A_{i-1}}Q_{B_{i-1}}}$, where $\mathcal{E}_i$ is a map from $E_{i-1}$ to $C_iE_i$ corresponding to Eve's box.}
\end{figure}

Since rDIQKD and DIQKD mainly differ in the quantum measurement phases $\mathcal{M}_i$ of the protocol, let us focus on this basic building block. 
At step $i$, and conditioning on the input classical variables $X_i,S_i,Z_i,Y_i$, we can describe this process as a CPTP map $\mathcal{M}_i:Q_{A_{i-1}}Q_{B_{i-1}}E_{i-1}\rightarrow A_iB_iC_iQ_{A_i}Q_{B_i}E_i$ that takes as input the quantum registers $Q_{A_{i-1}}$, $Q_{B_{i-1}}$, and $E_{i-1}$ of, respectively, Alice's private measurement device $\mathsf{A}$, Bob's private measurement device $\mathsf{B}$, and the eavesdropper Eve, and produces as output the classical variables $A_i$, $B_i$, $C_i$ along with updated quantum registers $Q_{A_{i}}$, $Q_{B_i}$, and $E_i$\footnote{Strictly speaking in any given round of the protocol, either the output $C_i$ is generated if $S_i=0$, or the output $B_i$ is generated if $S_i=1$. However, we can always assume that values are assigned both to $C_i$ and $B_i$ at each step $i$. For instance, we can set $C_i=\perp$ if $S_i=1$ and $B_i = \perp$ if $S_i=0$.}. 
This CTPT map has the structure outlined in Fig~\ref{fig:geat}. It is essentially similar to the corresponding map of a standard DIQKD protocol, except for the middle part under the control of Eve, which depends on additional random inputs $S_i$ and $Z_i$ and produces an additional outcome $C_i$.  However, this additional data is part of Eve's side information $E_i$ and thus does not affect the no-signaling condition necessary to apply the GEAT. More specifically, 
\begin{equation}
    \tr_{A_iB_iQ_{A_i}Q_{B_i}}\circ \mathcal{M}_i = \mathcal{E}_i\circ \tr_{Q_{A_{i-1}}Q_{B_{i-1}}}
\end{equation}
where this identity is easily visualized in Fig.~\ref{fig:geat-traced} and where $\mathcal{E}_i$ is a map from $E_{i-1}$ to $C_iE_i$. The map $\mathcal{M}_i$ is thus nonsignaling in the sense that tracing out Alice's and Bob's output quantum and classical registers $A_iB_iQ_{A_i}Q_{B_i}$, yields a map on Eve's systems that do not depend on the input systems $Q_{A_{i-1}}$ and $Q_{B_{i-1}}$ of Alice and Bob.
This no-signaling condition in the quantum measurement phase $\mathcal{M}_i$ of rDIQKD protocols allows the GEAT to be applied just in the same way as it would in DIQKD protocols and as outlined in Refs.~\cite{arnon2018practical,arnon2019simple}, ensuring that the no-signaling conditions of the maps $\mathcal{P}_i$ describing the intermediary EA protocol are satisfied.

\subsection{Asymptotic key-rate computation} \label{sec:keyrate_computation}
Consider a rDIQKD protocol where, according to the honest, ideal implementation, the source produces in each round the state $\rho_{AB}$ and the devices $\mathsf{A}$, $\mathsf{T}$, and $\mathsf{B}$ perform the POVMs $\mathsf{A}_x=\{\mathsf{A}_{a|x}\}_a$, $\mathsf{T}_z=\{\mathsf{T}_{c|z}\}_c$, and $\mathsf{B}_y=\{\mathsf{B}_{b|y}\}_b$, respectively, giving rise to the correlations $p(a,c|x,z)$ and $p(a,b|x,y)$ in Eq.~\eqref{eq:corr}.

The above discussion implies, as in the standard DIQKD setting, that the asymptotic key rate of such a rDIQKD protocol, assuming one-way public communication from Alice to Bob, is given by the IID Devetak-Winter rate \cite{Devetak2005}
\begin{equation} \label{eq:DWbound}
    r = H(K_A|XE) - H(K_A|K_B)\,,
\end{equation}
where  $H(K_A|XE)$ is the conditional von Neumann entropy of Alice's raw key symbol $K_A$ conditioned on the eavesdropper information $E$ and Alice's input $X$ and $H(K_A|K_B)$ is the Shannon entropy of Alice's raw key symbol $K_A$ conditioned on Bob's one $K_B$.  As the raw key is built from the outcomes of a subset $\mathcal{K}\subseteq\mathcal{X}\times\mathcal{Y}$ of all possible input pairs of Alice and Bob, these quantities can be expressed as the average values $H(K_A|XE) = \sum_{(x,y)\in\mathcal{K}} q(x,y) H(A_x|E)$ and $H(K_A|K_B) = \sum_{(x,y)\in\mathcal{K}} q(x,y) H(A_x|B_y)$ where $q(x,y)$ is the probability of choosing the input pair $(x,y)$ in a given key generation round. 

The term $H(K_A|K_B)$ captures the cost of error correction and can straightforwardly be computed from the distribution $p(a,b|x,y)$ fixed by the honest quantum strategy.
The term $H(K_A|XE)$ captures Eve's uncertainty about the measurement outcomes of Alice's measurements used for key generation. It is hard to compute because, in a DI setting where the devices are untrusted, it must be determined by taking the worst-case value over all possible quantum strategies compatible with the correlations $p(a,c|x,z)$ and $p(a,b|x,y)$ (which may differ from the honest strategy on the right-hand side of (\ref{eq:corr})). 

We briefly explain how $H(K_A|XE)$ can be lower bounded using NPO \cite{Navascues2007,Navascues2008,Pironio2010a} and the BFF method \cite{Brown2021}. We use this method in Sec.~\ref{sec:numerical} to provide numerical lower bounds on the key rate of various rDIQKD protocols of interest. The same techniques can be used to obtain min-trade-off functions and determine finite-size corrections to the Devetak-Winter rate.

\begin{figure}[t]
    \centering
    \begin{subfigure}{0.485\textwidth}
    \centering
    \includegraphics[width=\textwidth]{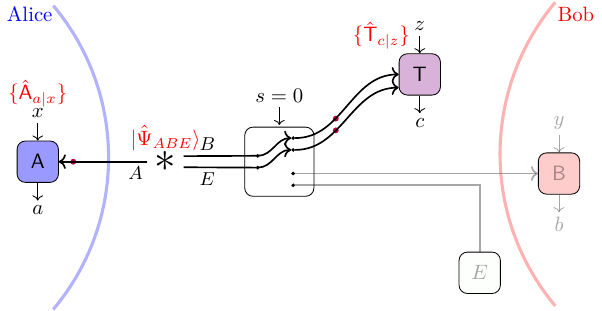}
    \caption{Case $s=0$.}
    \end{subfigure}
    \hfill
    \begin{subfigure}{0.485\textwidth}
    \centering
    \includegraphics[width=\textwidth]{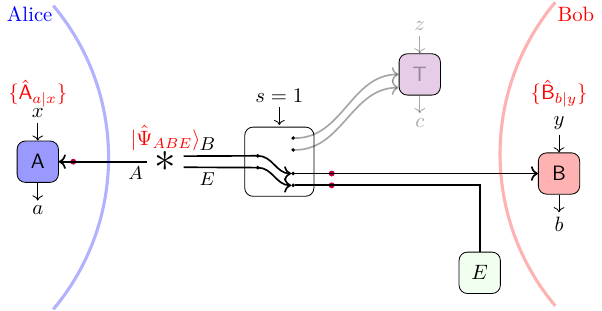}
    \caption{Case $s=1$.}
    \end{subfigure}
    \caption{Quantum model of Eve's strategies used to compute the single-round $H(K_A|XE)$ bound. \label{fig:qm-model} }
\end{figure}

We start by modeling the general behavior of the devices and Eve (in the setup of Fig.~\ref{fig:rdiqkd-setup-a}), which may differ from the honest implementation. This is depicted in Fig.~\ref{fig:qm-model}. The source first produces a state $\hat{\rho}_{ABE}$, where subsystem $A$ goes to Alice's device, subsystem $B$ will eventually go to Bob's device (if the switch setting is $s=1$) and subsystem $E$ characterizes Eve's initial quantum correlations with $A$ and $B$\footnote{Unlike the sequential depiction in Fig.~\ref{fig:geat}, there is no need to account for input and output quantum registers for either Alice or Bob, as we are focusing on a single round of an IID scenario. Furthermore, we address separately the cases where the switch input is $s=0$ and $s=1$. This simplifies the description of Eve's potential strategies.}. Subsystem $A$ is measured by device $\mathsf{A}$ through a measurement $\hat{\mathsf{A}}_x=\{\hat{\mathsf{A}}_{a|x}\}_a$. If $s=1$, subsystem $B$ is similarly measured by device $\mathsf{B}$ through a measurement $\hat{\mathsf{B}}_y=\{\hat{\mathsf{B}}_{b|y}\}$. If $s=0$, on the other hand, Eve, who holds the measurement device $\mathsf{T}$ performs a measurement $\hat{\mathsf{T}}_z=\{\hat{\mathsf{T}}_{c|z}\}_c$ that acts jointly on subsystems $B$ and $E$, as this is the most general thing she can do to simulate the honest correlations between $\mathsf{A}$ and $\mathsf{T}$. This potentially produces a postmeasurement state for Eve, but a description of this state is unnecessary for our purposes; since no key is extracted from a round where $s=0$, Eve's side information from such rounds is irrelevant to the analysis.

Without loss of generality, we can assume the initial state $\hat{\rho}_{ABE}$ to be a pure state $|\hat{\Psi}_{ABE}\rangle$, as any purifying system can be included in subsystem $E$. We can also assume that the measurements $\hat{\mathsf{A}}_x$, $\hat{\mathsf{B}}_y$ and $\hat{\mathsf{T}}_z$ are all projective, if necessary by considering enlarged systems. Finally, one might consider the possibility for Eve to do a joint operation on subsystems $B$ and $E$, depending on the value of the switch setting $s$, before proceeding as above. Without loss of generality, we can assume that these operations are unitary $\hat{U}^s_{BE}$, again by enlarging the Hilbert space if necessary. But then we can absorb the unitary $\hat{U}_{BE}^1$ in the definition of the initial state, $|\hat{\Psi}_{ABE}\rangle\leftarrow \hat{U}^1_{BE} |\hat{\Psi}_{ABE}\rangle$, and redefine the measurements at $\mathsf{T}$ as $\mathsf{T}_{c|z}\leftarrow \hat{U}^{1}_{BE} \hat{U}^{0 \dagger}_{BE} \mathsf{T}_{c|z} \hat{U}^{0 }_{BE} \hat{U}^{1\dagger}_{BE}$. There is thus actually no need to consider the operations $\hat{U}^s_{BE}$ explicitly in the modeling of the devices and Eve's strategies.

Altogether, the possible quantum strategies $\hat{\mathcal{Q}}=\{|\hat{\Psi}_{ABE}\rangle,\hat{\mathsf{A}}_{a|x},\hat{\mathsf{B}}_{b|y},\hat{\mathsf{T}}_{c|z}\}$ that Eve can use are fully characterized by the initial state $|\hat{\Psi}_{ABE}\rangle$, and the projective measurements $\{\hat{\mathsf{A}}_{a|x}\}_a$, $\{\hat{\mathsf{B}}_{b|y}\}_b$, and $\{\hat{\mathsf{T}}_{c|z}\}_c$. These should be constrained by the fact that they return the honest correlations $p(a,c|x,z)$ and $p(a,b|x,y)$:
\begin{eqnarray}
    p(a,c|x,z) &=& \langle \hat{\Psi}_{ABE}|\,\hat{\mathsf{A}}_{a|x} \otimes \hat{\mathsf{T}}_{c|z}\,|\hat{\Psi}_{ABE}\rangle\,,\label{eq:constr1}\\
    p(a,b|x,y) &=& \langle \hat{\Psi}_{ABE}|\,\hat{\mathsf{A}}_{a|x}\otimes \hat{\mathsf{B}}_{b|y}\otimes \mathbb{I}_E\,|\hat{\Psi}_{ABE}\rangle\,,\label{eq:constr2}
\end{eqnarray}
where we recall that $\hat{\mathsf{T}}_{c|z}$ acts jointly on subsystems $B$ and $E$.

To each strategy $\hat{\mathcal{Q}}$, we can associate the postmeasurement state $\sigma_{AXE}$
\begin{equation}
    \sigma_{AXE} = \sum_{ax} p(x) \ketbra{ax} \otimes \sigma_{E}^{a,x} \label{eq:post-meas} \, ,
\end{equation}
where 
\begin{equation}
    \sigma_E^{a,x} = \tr_{AB}(|\hat{\Psi}_{ABE}\rangle\langle \hat{\Psi}_{ABE}|\,(\hat{\mathsf{A}}_{a|x}\otimes \mathbb{I}_B \otimes \mathbb{I}_E)) \, ,
\end{equation}
is the unnormalized state held by Eve conditioned on Alice's input $x$ and output $a$. The conditional min-entropy can then be computed as 
\begin{equation}
    H(K_A|XE) = \inf_{\hat{\mathcal{Q}}|p} H(K_A|XE)_{\sigma_{AXE}} \,,\nonumber
\end{equation}
where the optimization runs over all quantum strategies $\hat{\mathcal{Q}}$ compatible with the honest correlations $p(a,c|x,z)$ and $p(a,b|x,y)$.

Notice now that the above optimization problem is almost identical to the optimization problem in a corresponding standard DIQKD protocol where Bob uses the input set $\mathcal{T}\times \mathcal{Y}$ and performs the measurements $\{\hat{\mathsf{T}}_z\}\times \{\hat{\mathsf{B}}_y\}$. The only difference with a regular DIQKD scenario is that the subset of measurements $\{\hat{\mathsf{T}}_z\}$ act on the joint systems $BE$, instead of just $B$. The BFF method \cite{Brown2021} based on NPO \cite{Navascues2007,Navascues2008,Pironio2010a} can then be used for lower bounding the conditional entropy $H(K_A|XE)$ in rDIQKD in almost the same way as in DIQKD. Since operators acting on different systems are replaced by commuting operators in NPO, the only difference with a standard BFF computation for DIQKD is that no commutation relations should be imposed between the $\mathsf{T}_z$ measurements and the BFF operators acting on Eve's system. We detail the NPO formulation corresponding to rDIQKD in Appendix~\ref{app:NCPOP}.

\section{Numerical results}
\label{sec:numerical}

Using the numerical technique discussed in the previous section, we now compute lower bounds on the key rate of several simple rDIQKD protocols.
We are interested in an rDIQKD setup in which the devices $\A$ and $\T$ have high detection efficiencies, i.e., are situated close to the source. For simplicity, we assign them identical (high) efficiencies, $\eta_{\A} = \eta_{\T} \equiv \eta_{\zs}$. Conversely, since we want to establish the key over long distances, Bob's device $\B$ must be situated far away and consequently has a lower efficiency, $\eta_{\B} \equiv \eta_{\zl} \ll  \eta_{\zs}$. For a given short-path efficiency $\eta_\zs$, we compute lower bounds on the asymptotic key rate $r$ as a function of the long-path detection efficiency $\eta_\zl$. 

In computing the key rate, we can use the Devetak-Winter formula corresponding to one-way communication from Alice to Bob, as in Eq.~\eqref{eq:DWbound}, or from Bob to Alice given by
$r = H(K_B|YE)-H(K_B|K_A)$. 
The entropy $H(K_B|YE)$ quantifies how well Eve can guess Bob's outcome. As Bob's device has a lower efficiency than Alice's, we expect that this will typically be easier for her than to guess Alice's outcome. Furthermore, since Bob holds a simple untrusted measurement device, Eve can apply to it the convex-combination attacks based on joint-measurability introduced in Ref.~\cite{masini2024one}, which imply particularly stringent limits on the minimal efficiency $\eta_{\zl}$ required to distill a secret key. We carried out numerical exploratory tests that confirm that the key rate is lower when computing it using one-way communication from Bob to Alice instead of the other direction. In the following, we thus compute all key rates using the bound (\ref{eq:DWbound}) holding for one-way communication from Alice to Bob.

Nondetection events $\emptyset$ corresponding to the situation where a detector fails to click can either be treated as separate measurement outcomes or binned with one of the other outcomes, say $\emptyset \mapsto +1$. This choice need not be the same in testing and key generation rounds and may also differ between the devices $\A$, $\T$ and $\B$. Binning the outcomes of Alice's device $\A$ in key generation rounds decreases the entropies $H(K_A|XE)$ and $H(K_A|K_B)$. However, numerical tests indicate that the decrease in $H(K_A|K_B)$ is more pronounced than for $H(K_A|XE)$ and thus the net effect on the key rate is positive. This is also what one observes in standard DIQKD protocols. Conversely, binning the outcomes of the device $\B$ in key generation rounds increases $H(K_A|K_B)$, as it decreases the information available to Bob for error correction, and thus lowers the key rate. In the following, we will therefore always bin the outcomes of $\A$  and keep the outcomes of $\B$ unbinned in key generation rounds, i.e., in computing the entropies $H(K_A|XE)$ and $H(K_A|K_B)$, the random variable $A$ corresponds to the binned version of Alice's outcome and the random variable $B$ to the unbinned version of Bob's outcome.

The bound on $H(K_A|XE)$ computed from the BFF method depends on the correlations $p(a,b|x,y)$ and $p(a,c|x,z)$ in testing rounds through the constraints (\ref{eq:constr1}) and (\ref{eq:constr2}). Binning the outcomes of the devices $\A$, $\T$, and $\B$ for testing rounds induces fewer restrictions on these correlations, yielding potentially lower values of $H(K_A|XE)$. Unfortunately, keeping no-click events separate in testing rounds substantially increases the size of the SDP relaxation problem used to compute $H(K_A|XE)$. For these reasons, we focus on protocols in which the outcomes of the devices $\A$ and $\T$ are always binned in testing rounds. For the outcomes of $\B$, we consider both situations where they are binned or not.  Although the latter scenario should yield higher key rates, binning the outcomes could enable us to implement SDP relaxation of the BFF NPO problem at a higher level, possibly resulting in improved key rates.

Besides detection efficiency, another important consideration in any experimental implementation is the impact of noise. We will consider a simple noise model, in which the state $\rho$ distributed by the source is mixed with white noise $\rho_{\rm noise} = \nu \rho + (1-\nu) \openone/4$, where ${\nu}$ is the visibility.

\subsection{A family of CHSH-BB84-type protocols}

We study a family of protocols where on Alice's side $\mathcal{X} = \{0,1\}$, $\mathcal{Z}=\{0,1\}$ and, in the honest implementation, the shared state is the two-qubit maximally entangled state $|\phi_+\rangle$ and the measurements are $\mathcal{X} = \{\sigma_z,\sigma_x\}$, $\mathcal{Z} = \{\frac{\sigma_z \pm \sigma_x}{\sqrt{2}}\}$. Thus the $\A/\T$ correlations are CHSH correlations and perfectly self-test the honest implementation when the efficiency $\eta_\zs=1$ and the visibility $\nu=1$. 

The protocols in our family only differ in the measurements of Bob and the subset of measurements used for key generation. We list the protocols we consider in Table~\ref{table:protocols}.
\begin{table*}[t] 
    \centering
    \begin{tabular}{lcccc}
    \hline
    Protocol               &$\mathcal{Y}$  &Ideal measurements &$\mathcal{K}$   & $y$ in SDP \\ \hline
    rCHSH      & $\{0,1,2\}$ & $ \{\sigma_z,\frac{\sigma_z \pm \sigma_x}{\sqrt{2}}\}$   & $\{(0,0)\}$  &  $\{1,2\}$        \\
    rBB84   & $\{0, 1\}$ & $\{\sigma_z,\sigma_x\}$ & \{(0,0)\} & \{0,1\}                               \\
    rCHSH-BB84 & $\{0,1,2,3\}$ & $\{\sigma_z,\sigma_x,\frac{\sigma_z \pm \sigma_x}{\sqrt{2}}\}$ & \{(0,0)\} & \{0,1,2,3\}                                         \\
    2-basis rBB84         & $\{0, 1\}$ & $\{\sigma_z,\sigma_x\}$ & \{(0,0),(1,1)\} & \{0,1\}                          \\
\hline
    \end{tabular}
    \caption{Family of routed rDIQKD protocols considered in this work. For each protocol, we indicate in each column: the inputs $\mathcal{Y}$ of Bob, the corresponding ideal measurements in the honest implementation, the settings $\mathcal{K}$ used for key generation, and finally which inputs are included in the SDP relaxation for $H(K_A|
    XE)$. \label{table:protocols}}
\end{table*}

The protocol rCHSH is the routed version of the standard DIQKD CHSH protocol introduced in Ref.~\cite{Acin2007}, the only difference being the added $\T$ measurements in the routed version. Bob has three inputs $\{0,1,2\}$ where the first one is used to establish the raw key shared with Alice and the rest are used to estimate the CHSH violation between Alice and Bob. To simplify the SDP relaxation used to compute $H(K_A|XE)$, we do not include the probabilities $p(ab|xy)$ corresponding to the inputs $y = 0$ in the NPO problem detailed in Appendix~\ref{app:NCPOP}.

The protocol rBB84 is the routed version of the standard BB84 protocol, in which the key is built from the $(x,y)=(0,0)$ inputs. This protocol is insecure in the DIQKD setting since the corresponding correlations admit a local model. However, the rBB84 correlations are LRQ for sufficiently high efficiencies $\eta_\zs$ and $\eta_\zl$, and visibility $\nu$ \cite{Lobo2023}. We will show that they lead to a positive key rate in the rDIQKD setting for a range of parameters. 
We also studied a two-basis version of the rBB84 protocol in which the key is built from the inputs $(x,y)\in \{(0,0), (1,1)\}$. While in principle, it could give an improvement over the one-basis version, the size of the corresponding SDP relaxation was too large to run it at the same level as the one-basis version and we did not observe this improvement. We therefore do not consider this protocol further in the following.

Finally, the rCHSH-BB84 protocol is a combination of the rCHSH and rBB84 protocols in which Bob performs the measurements appearing both in the rCHSH and rBB84 protocols. 

We note that the rDIQKD protocols that we analyze have not only a corresponding DIQKD version, obtained by removing the intermediate $\T$ measurements but also a semi-DIQKD prepare-and-measure analog. Indeed, the $\A/\T$ CHSH test effectively self-tests that Alice is remotely preparing the BB84 states when the efficiency $\eta_\zs=1$ and the visibility $\nu=1$. In this ideal case, all the above protocols are thus equivalent to a semi-DI prepare-and-measure (PM) version in which Alice's preparation is trusted to prepare the BB84 states and in which Bob's measurements are fully untrusted. In particular, the security of the rBB84 protocol in this ideal case is then equivalent to the security of the BB84 PM protocol in the one-sided DI setting \cite{mayers2001unconditional,berta2010uncertainty,woodhead2016semi,masini2024one}. However, the rBB84 protocol has the advantage that it is fully device independent, with the measurements performed at $\T$ certifying the BB84 preparations, while these preparations are instead assumed to be the correct ones in the one-sided DI PM version. When the efficiency $\eta_\zs<1$ and/or the visibility $\nu<1$, the rBB84 then can be thought of as a semi-DI BB84 protocol in which Alice's preparation is nonideal and only partly trusted. Similarly, the rCHSH protocol can be seen to be related to the semi-DI PM CHSH protocol considered in Ref.~\cite{woodhead2015secrecy} and the rCHSH-BB84 protocol to the semi-DI PM CHSH-BB84 protocol of Ref.~\cite{woodhead2012semi}.

 We computed bounds on the key rates for each of the protocols listed in Table~\ref{table:protocols} for different values of the short-path detection efficiency $\eta_\zs$ and the long-path detection efficiency $\eta_\zl$ and different visibilities $\nu$ using the BFF method and SDP relaxations. These results, as well as the codes used to generate them, are available at \cite{Git}. Below, we discuss several interesting aspects of the results.

 \subsection*{Binning vs not binning}
 \begin{figure*}
    \centering
    \begin{subfigure}{0.485\textwidth}
        \centering
        \includegraphics[width=\textwidth]{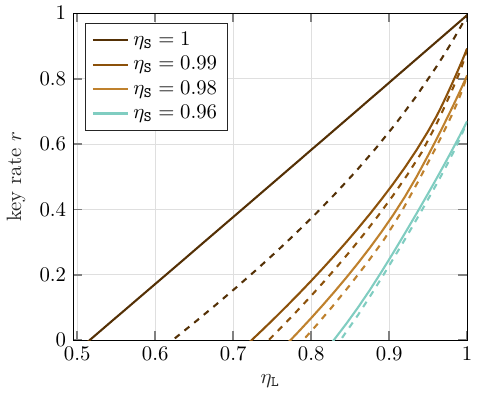}
         \caption{rCHSH protocol.}
        \end{subfigure}
        \hfill
        \begin{subfigure}{0.485\textwidth}
        \centering
        \includegraphics[width=\textwidth]{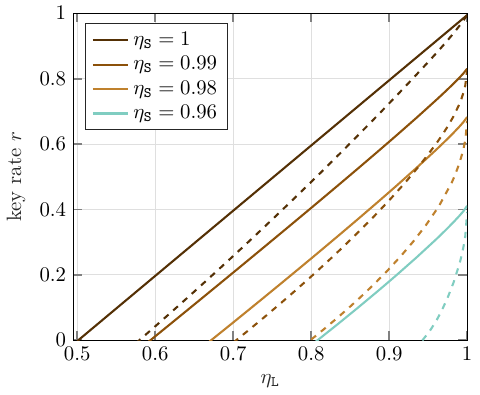}
        \caption{rBB84 protocol.\label{fig:binning_no_binning-b}}
        \end{subfigure}
    \caption{\emph{Binning vs not binning for $\B$.} Lower bounds on the asymptotic key rate $r$ as a function of $\eta_\zl$ for different values of $\eta_\zs$ with binning (dashed curves) and without binning (solid curves) and visibility $\nu =1$. Bounds are obtained at NPA level $2 + ABZ + AZB + AABZ + AAZB$.}
    \label{fig:binning_no_binning}
\end{figure*}

We first compare the key rates for every routed protocol with and without binning for the measurements at $\B$ in testing rounds (remember that we always view the no-click outcome $\emptyset$ of Bob as a separate outcome in key generation rounds and that we always bin the outcomes of $\A$ and $\T$). In every case, we find that the key rate is significantly better when Bob does not bin. This is consistent with the findings of Ref.~\cite{Lobo2023}, where it was shown that keeping no-click outcomes significantly decreases the critical efficiency at which correlations become SRQ. We plot the comparison in Fig.~\ref{fig:binning_no_binning} for the rCHSH and rBB84 protocols for visibility $\nu =1$.

\subsection*{Comparison to nonrouted protocols}
\begin{figure*}
    \centering
        \begin{subfigure}{\textwidth}
            \begin{subfigure}{0.32\textwidth}
                \centering
                \includegraphics[width=\textwidth]{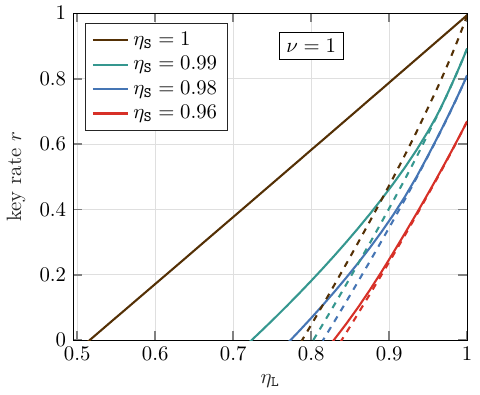}
            \end{subfigure}
            \begin{subfigure}{0.32\textwidth}
                \centering
                \includegraphics[width=\textwidth]{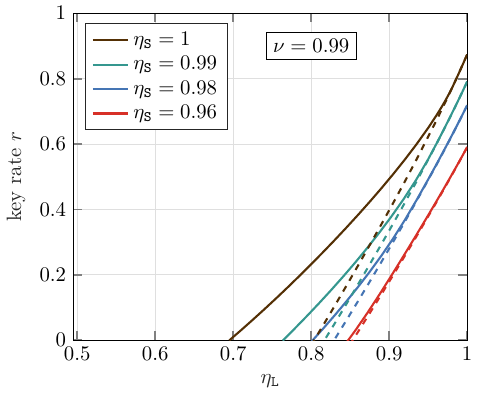}
            \end{subfigure}
            \begin{subfigure}{0.32\textwidth}
                \centering
                \includegraphics[width=\textwidth]{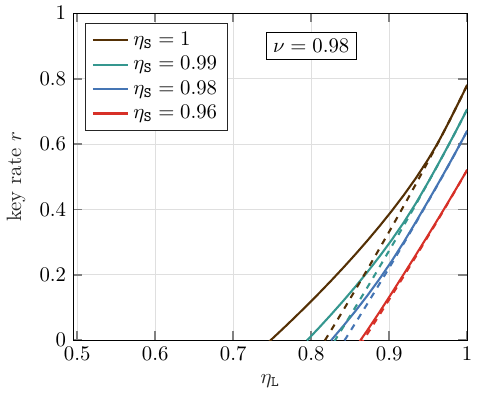}
            \end{subfigure}
        \addtocounter{subfigure}{0}
        \caption{CHSH vs rCHSH protocols.}
        \end{subfigure}
    \begin{subfigure}{\textwidth}
        \begin{subfigure}{0.32\textwidth}
            \centering
            \includegraphics[width=\textwidth]{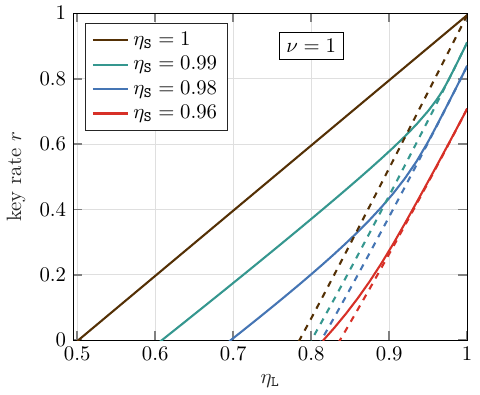}
        \end{subfigure}
        \begin{subfigure}{0.32\textwidth}
            \centering
            \includegraphics[width=\textwidth]{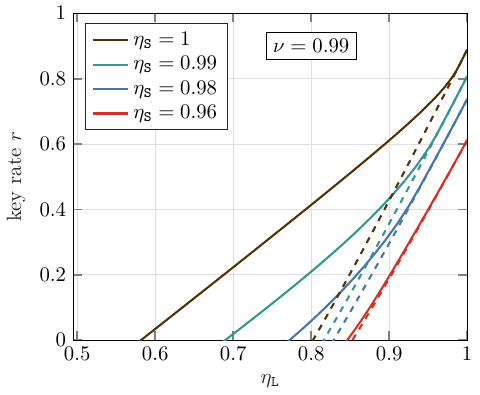}
        \end{subfigure}
        \begin{subfigure}{0.32\textwidth}
            \centering
            \includegraphics[width=\textwidth]{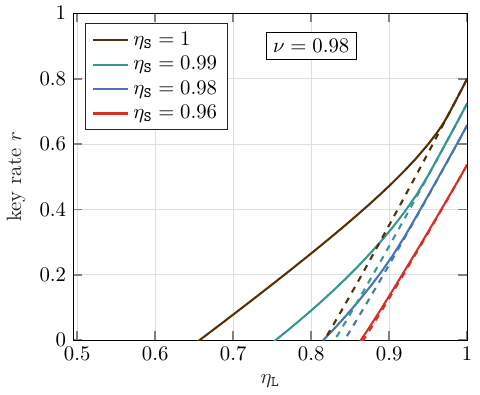}
        \end{subfigure}
    \addtocounter{subfigure}{0}
    \caption{CHSH-BB84 vs rCHSH-BB84 protocols.}
    \end{subfigure}
    \caption{\emph{Comparison to nonrouted DIQKD protocols.} Lower bounds on the asymptotic key rate $r$ for routed (solid curves) and non-routed (dashed curves) protocols when the outcomes of $\B$ are not binned. The bounds for CHSH and rCHSH protocols were obtained at NPA level $2 + ABZ + AZB + AABZ + AAZB$, whereas the bounds for CHSH-BB84 and rCHSH-BB84 were obtained at level $2 + ABZ + AZB$. \label{fig:routed_vs_standard}}
\end{figure*}

Let us now compare routed protocols to their nonrouted counterparts. There are two natural ways to do so. First, we can simply consider the effect of removing the device $\T$ from the routed protocols on the key rate. We plot the key rate of the rCHSH and rCHSH-BB84 protocols against their nonrouted versions in Fig. \ref{fig:routed_vs_standard}.  In both cases, we see that the introduction of a testing device $\T$ significantly improves the key rate. In particular, for $\nu = 1$, $\eta_\zs=0.99$, we obtain a critical detection efficiency of, respectively, $\eta_\zl \sim 0.72$ and $\eta_\zl \sim 0.61$ for the rCHSH and rCHSH-BB84 protocols, which represents an approximately $11\%$ and approximately $31\%$ improvement over their corresponding nonrouted protocols. 

To compare the corresponding improvement in the distance, let us write the detection efficiency $\eta$ as $\eta = \eta_f \eta_d$, where $1-\eta_d$ is the photon loss due to the transmission in the fiber, and $1-\eta_f$ is a fixed value that captures losses from all other sources. A fraction $1-\eta_d$ of the photons are absorbed in an optical fiber of transmission coefficient $\gamma$ dB km$^{-1}$ and length $-(1/\gamma) 10 \log_{10}(\eta_d)$ km. Assuming $\eta_f = 0.995$ for nonrouted protocols and for the device $\A$ in routed protocols, $\eta_f = 0.990$ for the devices $\T$ and $\B$ in routed protocols \footnote{We assume a lower value of $\eta_f$ for devices connected by a switch to take into account the additional losses from the switch.}, and $\gamma = 0.2$ dB km$^{-1}$, the total critical distance between $\A$ and $\B$ at which the key rate is positive is given by approximately $7.0$ km and approximately $10.6$ km for the rCHSH and rCHSH-rBB84 protocols, respectively. This represents an improvement of approximately $2.2$ km (approximately $45\%$), and approximately $5.8$ km (approximately $119$\%), respectively, over the nonrouted protocols.

However, this improvement declines rapidly for lower values of $\eta_\zs$ and $\nu$ due to the fact that the testing device $\T$ is less effective in certifying the correlations when the short-path test is of lower quality. For instance, at $\nu = 0.98$, $\eta_\zs=0.96$, the lower bounds on the key rates that we obtain for the standard and routed protocols are almost identical.

\begin{figure}
    \centering
   \includegraphics{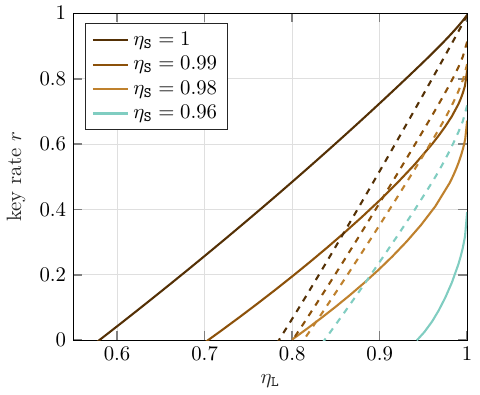}
    \caption{\emph{Comparison of the standard CHSH-BB84 protocol to rBB84.} Lower bounds on the asymptotic key rate $r$ as a function of $\eta_\zl$ for different values of $\eta_\zs$ for rBB84 and CHSH-BB84 with binning of the outcomes of $\B$ and visibility $\nu = 1$. Bounds are obtained at NPA level $2 + ABZ + AZB + AABZ + AAZB$. \label{fig:rBB84_vs_CHSH-BB84}}
\end{figure}

The above comparison between DIQKD and rDIQKD protocols is not meaningful for the rBB84 protocol, since BB84 correlations are local without the testing device $\T$ and thus cannot be used for DIQDK.
A second way to compare rDIQKD to DIQKD protocols is to view the routed protocols as a DIQKD protocol in which some of the measurements of Bob used for testing rounds have been moved closer to the source, to the $\T$ device. The rBB84 protocol can be seen in this way as originating from a standard CHSH-BB84 DIQKD protocol in which the remote $\frac{\sigma_z\pm\sigma_x}{\sqrt{2}}$ measurements of Bob are moved to the $\T$ device.  While this increases the quality of these measurements due to improved detection efficiency at shorter distances, it comes at the cost of revealing to Eve that certain rounds are only used for testing. We clearly see this trade-off in Fig.~\ref{fig:rBB84_vs_CHSH-BB84}, where we have assumed $\nu = 1$ for simplicity. For high values of $\eta_{\zs}$ and low values of $\eta_{\zl}$, the routed protocol performs better due to the advantage gained in certifying Alice's devices. But as $\eta_{\zs}$ decreases and $\eta_{\zl}$ increases, the nonrouted protocol improves and eventually outperforms the routed one, as it reveals less information to Eve about which measurements Bob's device is performing. At $\eta_{\zs} = 0.96$, the nonrouted protocol outperforms the routed protocol for all values of $\eta_{\zl}$, while at $\eta_{\zs} = 1$, the routed protocol outperforms the nonrouted protocol for all values of $\eta_{\zl}$.


\subsection*{Comparison of different protocols}
\begin{figure*}
    \centering
    \begin{subfigure}{0.49\textwidth}
    \centering
    \includegraphics[scale=0.95]{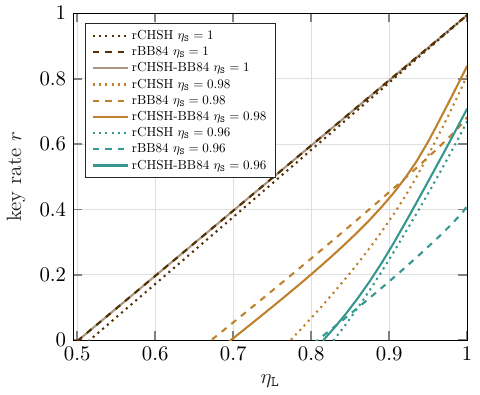}
    \caption{Plot for different values of $\eta_\zs$ and visibility $\nu = 1$.\label{fig:protocol_comparison-a}}
    \end{subfigure}
    \hfill
    \begin{subfigure}{0.49\textwidth}
    \centering
    \includegraphics[scale=0.95]{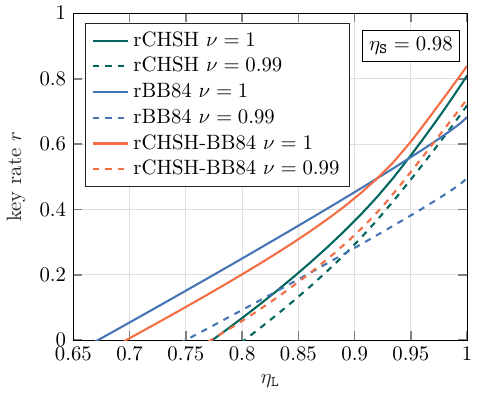}
    \caption{Plot for $\eta_\zs = 0.98$ and visibility $\nu = 1$ and $\nu = 0.99$.\label{fig:protocol_comparison-b}}
    \end{subfigure}
    \caption{\emph{Comparison of different protocols.} Lower bounds on the asymptotic key rate $r$ as a function of $\eta_\zl$  for the different routed protocols in Table~\ref{table:protocols} in the case where Bob does not bin. Bounds for rCHSH-BB84 (rCHSH and rBB84) are obtained at NPA level $2 + ABZ + AZB (+ AABZ + AAZB)$.}
    \label{fig:protocol_comparison}
\end{figure*}

We compare in Fig.~\ref{fig:protocol_comparison} the key rates for the different routed protocols listed in Table~\ref{table:protocols}, in the case where the outcomes of $\B$ are not binned. For high-quality short-range tests, the rBB84 protocol performs better than the rCHSH protocol, but this performance decreases more rapidly with $\eta_\zs$. The best performance should be obtained from the rCHSH-BB84 protocol since the key rates for this protocol are in principle at least as good as the ones obtained for the rCHSH and rBB84 protocols. Though we do find that the rCHSH-BB84 protocol always outperforms the rCHSH protocol, we see that there exist regions of the parameter (when $\eta_\zl$ is low), where our rBB84 bounds are higher than the rCHSH-BB84 bounds. This is due to the fact that the rCHSH-BB84 bounds were computed at a lower level of the NPA hierarchy than the other protocols due to the bigger SDP size.  We thus expect that the key rates for the rCHSH-BB84 protocol could be greatly improved by computing them at a higher level of the NPA hierarchy.

We showcase the effect of finite visibility $\nu$ in Fig.~\ref{fig:protocol_comparison-b} for $\eta_\zs = 0.98$. We see that the key rate is very sensitive to finite visibility, but the effect is more pronounced for the rBB84 protocol. This is consistent with our previous observations: the rBB84 protocol performs best when the $\A/\T$ test is of high quality ($\nu=1$), while the rCHSH protocol is more robust to noise ($\nu=0.99$).

\subsection*{Additional remarks}
We finish with several additional remarks on the results obtained.
First, as we explained earlier, in the limiting case where $\eta_\zs = 1$, the $\A/\T$ correlations effectively self-test the BB84 preparations. The rBB84 protocol is then equivalent to the one-sided PM BB84 protocol where Bob's measurement is fully untrusted.
If the only information used in parameter estimation, assuming Bob bins his outcomes, is the standard quantum bit-error rate (QBER) $q_x$ in the $\sigma_x$ basis (the $\sigma_z$ basis is used for key generation) then the asymptotic key rate is given by the Shor-Preskill rate \cite{Shor2000,woodhead2016semi}  $r = 1- h(q_x)-H(A_0|B_0)$, where $h(\cdot)$ is the binary entropy. A simple calculation gives $q_x = \frac{1-\eta_\zl}{2}$ and $H(A_0|B_0) = 1-\eta_\zl$, implying $r = 1- h\qty(\frac{1-\eta_{\zl}}{2})-(1-\eta_\zl)$. We verified numerically that we recover this rate if we impose in the BBF method, besides the ideal CHSH $\A/\T$ correlations, only the QBER value for the $\A/\B$ correlations. We then find that the key rate is positive as long as $\eta_\zl\gtrsim 65.9\%$.

Interestingly, when we compute the key rate by fixing the full set of $\A/\B$ correlations and not only the QBER in the $\sigma_x$ basis, we find as illustrated in Fig.~\ref{fig:binning_no_binning-b} that for $\eta_\zs=1$, the key rate in the rBB84 protocol is positive for $\eta_\zl\gtrsim 58.5\%$ when binning $\B$ outcomes and $\eta_\zl\gtrsim 50\%$ when nonbinning $\B$ outcomes. These results, due to the equivalence mentioned above, also apply to one-sided PM BB84 protocols and imply that the Shor-Preskill rate can be significantly improved by taking into account the full set of $\A/\B$ correlations. This is consistent with the recent results of Ref.~\cite{masini2024one}. 

We also observe that the thresholds of $\eta_\zl\gtrsim 58.5\%$ (binning) and $\eta_\zl\gtrsim 50\%$ (nonbinning) at which the key rate of the rBB84 vanished correspond precisely to the point where the rBB84 correlations become SQR \cite{Lobo2023}. This is to be contrasted with DIQKD protocols in which the key rate typically vanishes well before the underlying correlations become local.

Finally, the $\eta_\zl\gtrsim 50\%$ threshold for rBB84 obtained in the case of nonbinning is optimal according to the upper bound (\ref{eq:bound-model}) and the general attacks of Ref.~\cite{masini2024joint}. Although the same upper bound applies to DIQKD protocols, it is not clear if it can be achieved with a simple qubit-based protocol as in the routed case.

\section{Conclusion and outlook}
\label{sec:conclusion}
We considered routed DIQKD protocols based on the recently introduced routed Bell configuration \cite{Lobo2023,Chaturvedi2022}. Our work is motivated by the finding that introducing an additional measurement device close to the source, allowing for improved testing of quantum correlations at short distance, effectively lowers the visibility and detection efficiency requirements for certifying long-range quantum correlations \cite{Lobo2023}, which are a prerequisite for DI security.

We outlined how the GEAT applies to rDIQKD protocols, essentially in the same way the EAT applies to DIQKD protocols, enabling reducing the security analysis to a single round. We then applied the Brown-Fawzi-Fawzi method \cite{Brown2021} combined with semidefinite relaxation hierarchies \cite{Navascues2007,Navascues2008,Pironio2010a} to numerically compute lower bounds on the asymptotic key rate for several routed DIQKD protocols.

We focused on a simple family of two-qubit rDIQKD protocols, in which Alice's device $\A$ and Bob's device $\B$ establish CHSH correlations, BB84 correlations, or a combination of both. In all cases, the additional short-range testing device $\T$ is used to generate CHSH correlations with $\A$, thus enabling in the ideal case a perfect self-test of the source and Alice's measurements.
Consistent with the previous study on quantum correlations in routed Bell scenarios \cite{Lobo2023}, we find that for low-noise $\A/\T$ correlations (i.e., high detection efficiencies at short distance), rDIQKD protocols based on BB84 correlations between $\A$ and $\B$, whose nonlocality can be certified in routed setups, tolerate the lowest detection efficiency for the long-range device $\B$. When the short-range devices are ideal, rBB84 protocols can be secure up to the theoretical limit \cite{Lobo2023,masini2024joint} of $\eta_\zl\gtrsim 50\%$ for the long-range device $\B$. 
On the other hand, rCHSH protocols have absolute higher detection efficiency requirements for $\B$, but they are more robust to imperfections in the $\A/\T$ correlations. A combined CHSH-BB84 protocol is expected to perform best in both regimes, though clearly establishing this would likely require computing the key rates at a higher level of the SDP hierarchy.

While the advantages of routed protocols are significant, enabling the lowering of the detection efficiency of Bob's device $\B$ up to approximately $30\%$  compared to nonrouted protocols, they crucially depend on the ability to generate high-quality short-range correlations between Alice's device $\A$ and the additional testing device $\T$. This is further complicated by the need for a switch to implement routing to either the testing device $\T$ or Bob's device $\B$, which can be actively controlled at a sufficiently high rate, leading to additional noise and inefficiencies. Nevertheless, given that routed protocols can be implemented modularly in existing DIQKD setups at relatively minimal experimental overhead, they may facilitate the prospects of a full photonic implementation of DIQKD. 

Our work suggests several interesting directions for future work. First, it would be interesting to study more general rDQIKD protocols, particularly with more inputs and outputs, to overcome the ultimate limitations implied by convex-combination attacks based on joint measurability \cite{masini2024joint}. Second, it would be interesting to consider more complicated topologies, featuring multiple switches and testing devices, which could potentially provide more robust certification of the device $\B$. Along this line, we note that the DIQKD protocol introduced in Ref.~\cite{lim2013device} can be seen as a routed protocol with an additional Bell-state measurement between Alice and Bob to further limit the effect of losses in the channel. The numerical methods used in this work could possibly be improved to provide better bounds on the key rate of this protocol or study variants of it.
Finally, it would be interesting to explicitly model imperfections in the switch and explore the experimental feasibility of routed DIQKD protocols with current technology.

On a more conceptual level, the link between rDIQKD protocols and one-sided PM protocols is particularly interesting and deserves to be further explored. In particular, we note that the threshold $\eta_\zl\gtrsim 50\%$ for the rBB84 protocol when the $\A/\T$ correlations perfectly self-test the BB84 preparations imply that the traditional BB84 protocol can be secure up to this threshold in a one-sided setting where Bob's device is fully untrusted \cite{mayers2001unconditional,berta2010uncertainty,woodhead2016semi}. This finding, consistent with a similar observation in Ref.~\cite{masini2024one}, shows that the Shor-Preskill rate for the one-sided PM BB84 protocol can be significantly improved by taking into account the full set of $\A/\B$ correlations and keeping nondetection events as a separate outcome.

\emph{Note added.} While completing this work, we were made aware of similar work by E. Tan and R. Wolf  \cite{Tan2024}.

\begin{acknowledgments}
S.P. acknowledges funding from the VERIqTAS project within the QuantERA II Programme that has received funding from the European Union's Horizon 2020 research and innovation program under Grant Agreement No 101017733 and the F.R.S-FNRS Pint-Multi program under Grant Agreement R.8014.21, from the European Union's Horizon Europe research and innovation program under the project ``Quantum Security Networks Partnership'' (QSNP, grant agreement No 101114043),  from the F.R.S-FNRS through the PDR T.0171.22, from the FWO and F.R.S.-FNRS under the Excellence of Science (EOS) program project 40007526, from the FWO through the BeQuNet SBO project S008323N, from the Belgian Federal Science Policy through the contract RT/22/BE-QCI and the EU ``BE-QCI'' program.
S.P. is a Research Director of the Fonds de la Recherche Scientifique - FNRS. E.P.L. acknowledges support from the Fonds de la Recherche Scientifique - FNRS through a FRIA grant. J.P. acknowledges support from NCCR-SwissMAP. T.L.D. benefited from a government grant managed by the Agence Nationale de la Recherche under the Plan France 2030 with the reference ANR-22-PETQ-0009. This project was funded within the QuantERA II Programme which has received funding from the European Union's Horizon 2020 research and innovation program under Grant Agreement No 101017733.
\end{acknowledgments}

Funded by the European Union. Views and opinions expressed are however those of the authors only and do not necessarily reflect those of the European Union. The European Union cannot be held responsible for them.\\

\emph{Author contributions}. Everyone contributed equally in formulating the ideas and obtaining the main results. The codes for obtaining the numerical results were written by T.L.R-D and E.P.L. The codes were run and results plotted by T.L.R-D, E.P.L, and J.P. E.P.L, J.P, and S.P wrote the manuscript. The figures were made by E.P.L with inputs from S.P and J.P. S.P supervised the project. The authors are listed alphabetically.

\appendix
\section{Explicit formulation of BFF-NPO for lower bounding the conditional entropy $H(A_x|E)$} \label{app:NCPOP}

The Brown-Fawzi-Fawzi (BFF) method \cite{Brown2021a} provides a hierarchy of successively tighter approximations to the von Neumann entropy. As explained in the main text, we can use it in the context of rDIQKD in the same way it is used in DIQDK, modulo the introduction of operators $\hat{\T}_{c|z}$ that act on the joint systems $BE$, and hence do not commute with the BFF operators $Z_a$ acting on Eve's system $E$. Explicitly, the BFF-NPO formulation for lower bounding the conditional entropy $H(A_x|E)$ is as follows.

 Let $V_i$ be the solutions to the following NPO problems, where $t_i$ and $w_i$ are the nodes and weights of an $m$-point Gauss-Radau quadrature on $[0,1]$ with an endpoint at $t_m = 1$, and $\alpha_i = \frac{3}{2} \max\{\frac{1}{t_i}, \frac{1}{1-t_i}\}$:
\begin{align*}
V_i =\inf&\sum_a \bra{\psi} \A_{a|x^\ast}(Z_{a } +Z_{a }^\dagger+(1-t_i)Z_{a }^\dagger Z_{a })+ t_iZ_{a }Z_{a }^\dagger\ket{\psi}\hfill \,\\
\text{s.t.}&  \bra{\psi} \A_{a|x} \B_{b|y} \ket{\psi} = p(a,b|x,y), \quad \forall a,b,x,y, \\
&\bra{\psi} \A_{a|x} \T_{c|z} \ket{\psi} = p(a,c|x,z), \quad \forall a,c,x,z,\\
&\A_{a|x}\A_{a'|x} = \delta_{aa'}\A_{a|x}, \quad \forall a,a',x,\\
& \B_{b|y}\B_{b'|y} = \delta_{bb'}\B_{b|y}, \quad \forall b,b',y,\\
& \T_{c|z}\T_{c'|z} = \delta_{cc'}\T_{c|z}, \quad \forall c,c',z,\\
& [\A_{a|x},\B_{b|y}] = [\A_{a|x},\T_{c|z}] =  0, \quad\forall a,b,c,x,y,z, \\
& [\A_{a|x},Z_{a' }^{(\dagger)}] =[\B_{b|y},Z_{a }^{(\dagger)}]= 0,\quad\forall a,a',b ,x,y, \\
& Z_{a }^\dagger Z_{a } \leq \alpha_i, \quad\forall a, \\
& Z_{a } Z_{a }^\dagger \leq \alpha_i,\quad\forall a,,
\end{align*} 
where the minimization is over the state $|\psi\rangle$ and the Hermitian operators $\A_{a|x}$, $\B_{b|y}$, $\T_{c|z}$, and non-Hermitian operators $Z_a$.
The von Neumann entropy is then lower bounded by \cite{Brown2021a},
\begin{align*}
    H(A|E,X=x^\ast) \ge    &\sum_{i=1}^{m-1} \frac{w_i}{t_i \ln 2} (1+V_i) \, .
\end{align*}
We can compute a lower bound on the NPO optimums $V_i$, hence on $H(A_x|E)$, by relaxing them to semidefinite programs \cite{Navascues2007,Navascues2008,Pironio2010a}. The numerical results presented in this work have been computed using $m = 12$ points in Gauss-Radau quadrature and SDP relaxations corresponding to level 2 and some additional monomials, as detailed in the main text. Note that, as explained in Sec.~\ref{sec:keyrate_computation}, we will think of the input sets $\mathcal{T}\times\mathcal{Y}$ as belonging to Bob while computing key rates. This means, for example, that level `$ABZ$' consists of monomials of the form $LMN$ where $L \in \{\A_{a|x}\}_{a,x} \,,\, M \in \{B_{b|y}\}_{b,y} \bigcup \, \{T_{c|z}\}_{c,z} \,,$ and $ N \in \{Z_{a}\}_a $. For standard DIQKD protocols $\{T_{c|z}\}_{c,z}$ is empty.

\bibliography{refs.bib}

\begin{thebibliography}{41}%
\makeatletter
\providecommand \@ifxundefined [1]{%
 \@ifx{#1\undefined}
}%
\providecommand \@ifnum [1]{%
 \ifnum #1\expandafter \@firstoftwo
 \else \expandafter \@secondoftwo
 \fi
}%
\providecommand \@ifx [1]{%
 \ifx #1\expandafter \@firstoftwo
 \else \expandafter \@secondoftwo
 \fi
}%
\providecommand \natexlab [1]{#1}%
\providecommand \enquote  [1]{``#1''}%
\providecommand \bibnamefont  [1]{#1}%
\providecommand \bibfnamefont [1]{#1}%
\providecommand \citenamefont [1]{#1}%
\providecommand \href@noop [0]{\@secondoftwo}%
\providecommand \href [0]{\begingroup \@sanitize@url \@href}%
\providecommand \@href[1]{\@@startlink{#1}\@@href}%
\providecommand \@@href[1]{\endgroup#1\@@endlink}%
\providecommand \@sanitize@url [0]{\catcode `\\12\catcode `\$12\catcode `\&12\catcode `\#12\catcode `\^12\catcode `\_12\catcode `\%12\relax}%
\providecommand \@@startlink[1]{}%
\providecommand \@@endlink[0]{}%
\providecommand \url  [0]{\begingroup\@sanitize@url \@url }%
\providecommand \@url [1]{\endgroup\@href {#1}{\urlprefix }}%
\providecommand \urlprefix  [0]{URL }%
\providecommand \Eprint [0]{\href }%
\providecommand \doibase [0]{https://doi.org/}%
\providecommand \selectlanguage [0]{\@gobble}%
\providecommand \bibinfo  [0]{\@secondoftwo}%
\providecommand \bibfield  [0]{\@secondoftwo}%
\providecommand \translation [1]{[#1]}%
\providecommand \BibitemOpen [0]{}%
\providecommand \bibitemStop [0]{}%
\providecommand \bibitemNoStop [0]{.\EOS\space}%
\providecommand \EOS [0]{\spacefactor3000\relax}%
\providecommand \BibitemShut  [1]{\csname bibitem#1\endcsname}%
\let\auto@bib@innerbib\@empty
\bibitem [{\citenamefont {Ac\'{\i}n}\ \emph {et~al.}(2007)\citenamefont {Ac\'{\i}n}, \citenamefont {Brunner}, \citenamefont {Gisin}, \citenamefont {Massar}, \citenamefont {Pironio},\ and\ \citenamefont {Scarani}}]{Acin2007}%
  \BibitemOpen
  \bibfield  {author} {\bibinfo {author} {\bibfnamefont {A.}~\bibnamefont {Ac\'{\i}n}}, \bibinfo {author} {\bibfnamefont {N.}~\bibnamefont {Brunner}}, \bibinfo {author} {\bibfnamefont {N.}~\bibnamefont {Gisin}}, \bibinfo {author} {\bibfnamefont {S.}~\bibnamefont {Massar}}, \bibinfo {author} {\bibfnamefont {S.}~\bibnamefont {Pironio}},\ and\ \bibinfo {author} {\bibfnamefont {V.}~\bibnamefont {Scarani}},\ }\bibfield  {title} {\bibinfo {title} {Device-independent security of quantum cryptography against collective attacks},\ }\href {https://doi.org/10.1103/physrevlett.98.230501} {\bibfield  {journal} {\bibinfo  {journal} {Physical Review Letters}\ }\textbf {\bibinfo {volume} {98}},\ \bibinfo {pages} {230501} (\bibinfo {year} {2007})},\ \Eprint {https://arxiv.org/abs/quant-ph/0702152} {arXiv:quant-ph/0702152} \BibitemShut {NoStop}%
\bibitem [{\citenamefont {Brunner}\ \emph {et~al.}(2014)\citenamefont {Brunner}, \citenamefont {Cavalcanti}, \citenamefont {Pironio}, \citenamefont {Scarani},\ and\ \citenamefont {Wehner}}]{Brunner2014}%
  \BibitemOpen
  \bibfield  {author} {\bibinfo {author} {\bibfnamefont {N.}~\bibnamefont {Brunner}}, \bibinfo {author} {\bibfnamefont {D.}~\bibnamefont {Cavalcanti}}, \bibinfo {author} {\bibfnamefont {S.}~\bibnamefont {Pironio}}, \bibinfo {author} {\bibfnamefont {V.}~\bibnamefont {Scarani}},\ and\ \bibinfo {author} {\bibfnamefont {S.}~\bibnamefont {Wehner}},\ }\bibfield  {title} {\bibinfo {title} {{B}ell nonlocality},\ }\href {https://doi.org/10.1103/revmodphys.86.419} {\bibfield  {journal} {\bibinfo  {journal} {Review of Modern Physics}\ }\textbf {\bibinfo {volume} {86}},\ \bibinfo {pages} {419} (\bibinfo {year} {2014})},\ \Eprint {https://arxiv.org/abs/1303.2849} {arXiv:1303.2849 [quant-ph]} \BibitemShut {NoStop}%
\bibitem [{\citenamefont {Pearle}(1970)}]{Pearle1970}%
  \BibitemOpen
  \bibfield  {author} {\bibinfo {author} {\bibfnamefont {P.~M.}\ \bibnamefont {Pearle}},\ }\bibfield  {title} {\bibinfo {title} {Hidden-variable example based upon data rejection},\ }\href {https://doi.org/10.1103/PhysRevD.2.1418} {\bibfield  {journal} {\bibinfo  {journal} {Physical Review D}\ }\textbf {\bibinfo {volume} {2}},\ \bibinfo {pages} {1418} (\bibinfo {year} {1970})}\BibitemShut {NoStop}%
\bibitem [{\citenamefont {Clauser}\ and\ \citenamefont {Horne}(1974)}]{Clauser1974}%
  \BibitemOpen
  \bibfield  {author} {\bibinfo {author} {\bibfnamefont {J.~F.}\ \bibnamefont {Clauser}}\ and\ \bibinfo {author} {\bibfnamefont {M.~A.}\ \bibnamefont {Horne}},\ }\bibfield  {title} {\bibinfo {title} {Experimental consequences of objective local theories},\ }\href {https://doi.org/10.1103/PhysRevD.10.526} {\bibfield  {journal} {\bibinfo  {journal} {Physical Review D}\ }\textbf {\bibinfo {volume} {10}},\ \bibinfo {pages} {526} (\bibinfo {year} {1974})}\BibitemShut {NoStop}%
\bibitem [{\citenamefont {Garg}\ and\ \citenamefont {Mermin}(1987)}]{Garg1987}%
  \BibitemOpen
  \bibfield  {author} {\bibinfo {author} {\bibfnamefont {A.}~\bibnamefont {Garg}}\ and\ \bibinfo {author} {\bibfnamefont {N.~D.}\ \bibnamefont {Mermin}},\ }\bibfield  {title} {\bibinfo {title} {Detector inefficiencies in the {Einstein-Podolsky-Rosen} experiment},\ }\href {https://doi.org/10.1103/PhysRevD.35.3831} {\bibfield  {journal} {\bibinfo  {journal} {Physical Review D}\ }\textbf {\bibinfo {volume} {35}},\ \bibinfo {pages} {3831} (\bibinfo {year} {1987})}\BibitemShut {NoStop}%
\bibitem [{\citenamefont {Shalm}\ \emph {et~al.}(2021)\citenamefont {Shalm}, \citenamefont {Zhang}, \citenamefont {Bienfang}, \citenamefont {Schlager}, \citenamefont {Stevens}, \citenamefont {Mazurek}, \citenamefont {Abellán}, \citenamefont {Amaya}, \citenamefont {Mitchell}, \citenamefont {Alhejji}, \citenamefont {Fu}, \citenamefont {Ornstein}, \citenamefont {Mirin}, \citenamefont {Nam},\ and\ \citenamefont {Knill}}]{Shalm2021}%
  \BibitemOpen
  \bibfield  {author} {\bibinfo {author} {\bibfnamefont {L.~K.}\ \bibnamefont {Shalm}}, \bibinfo {author} {\bibfnamefont {Y.}~\bibnamefont {Zhang}}, \bibinfo {author} {\bibfnamefont {J.~C.}\ \bibnamefont {Bienfang}}, \bibinfo {author} {\bibfnamefont {C.}~\bibnamefont {Schlager}}, \bibinfo {author} {\bibfnamefont {M.~J.}\ \bibnamefont {Stevens}}, \bibinfo {author} {\bibfnamefont {M.~D.}\ \bibnamefont {Mazurek}}, \bibinfo {author} {\bibfnamefont {C.}~\bibnamefont {Abellán}}, \bibinfo {author} {\bibfnamefont {W.}~\bibnamefont {Amaya}}, \bibinfo {author} {\bibfnamefont {M.~W.}\ \bibnamefont {Mitchell}}, \bibinfo {author} {\bibfnamefont {M.~A.}\ \bibnamefont {Alhejji}}, \bibinfo {author} {\bibfnamefont {H.}~\bibnamefont {Fu}}, \bibinfo {author} {\bibfnamefont {J.}~\bibnamefont {Ornstein}}, \bibinfo {author} {\bibfnamefont {R.~P.}\ \bibnamefont {Mirin}}, \bibinfo {author} {\bibfnamefont {S.~W.}\ \bibnamefont {Nam}},\ and\ \bibinfo {author} {\bibfnamefont {E.}~\bibnamefont {Knill}},\ }\bibfield  {title} {\bibinfo {title} {Device-independent randomness expansion with entangled photons},\ }\href {https://doi.org/10.1038/s41567-020-01153-4} {\bibfield  {journal} {\bibinfo  {journal} {Nature Physics}\ }\textbf {\bibinfo {volume} {17}},\ \bibinfo {pages} {452} (\bibinfo {year} {2021})},\ \Eprint {https://arxiv.org/abs/1912.11158} {arXiv:1912.11158 [quant-ph]} \BibitemShut {NoStop}%
\bibitem [{\citenamefont {Li}\ \emph {et~al.}(2021)\citenamefont {Li}, \citenamefont {Zhang}, \citenamefont {Liu}, \citenamefont {Zhao}, \citenamefont {Bai}, \citenamefont {Liu}, \citenamefont {Zhao}, \citenamefont {Peng}, \citenamefont {Zhang}, \citenamefont {Zhang}, \citenamefont {Munro}, \citenamefont {Ma}, \citenamefont {Zhang}, \citenamefont {Fan},\ and\ \citenamefont {Pan}}]{Li2021}%
  \BibitemOpen
  \bibfield  {author} {\bibinfo {author} {\bibfnamefont {M.-H.}\ \bibnamefont {Li}}, \bibinfo {author} {\bibfnamefont {X.}~\bibnamefont {Zhang}}, \bibinfo {author} {\bibfnamefont {W.-Z.}\ \bibnamefont {Liu}}, \bibinfo {author} {\bibfnamefont {S.-R.}\ \bibnamefont {Zhao}}, \bibinfo {author} {\bibfnamefont {B.}~\bibnamefont {Bai}}, \bibinfo {author} {\bibfnamefont {Y.}~\bibnamefont {Liu}}, \bibinfo {author} {\bibfnamefont {Q.}~\bibnamefont {Zhao}}, \bibinfo {author} {\bibfnamefont {Y.}~\bibnamefont {Peng}}, \bibinfo {author} {\bibfnamefont {J.}~\bibnamefont {Zhang}}, \bibinfo {author} {\bibfnamefont {Y.}~\bibnamefont {Zhang}}, \bibinfo {author} {\bibfnamefont {W.~J.}\ \bibnamefont {Munro}}, \bibinfo {author} {\bibfnamefont {X.}~\bibnamefont {Ma}}, \bibinfo {author} {\bibfnamefont {Q.}~\bibnamefont {Zhang}}, \bibinfo {author} {\bibfnamefont {J.}~\bibnamefont {Fan}},\ and\ \bibinfo {author} {\bibfnamefont {J.-W.}\ \bibnamefont {Pan}},\ }\bibfield  {title} {\bibinfo {title} {Experimental realization of device-independent quantum randomness expansion},\ }\href {https://doi.org/10.1103/physrevlett.126.050503} {\bibfield  {journal} {\bibinfo  {journal} {Physical Review Letters}\ }\textbf {\bibinfo {volume} {126}},\ \bibinfo {pages} {050503} (\bibinfo {year} {2021})},\ \Eprint {https://arxiv.org/abs/1902.07529} {arXiv:1902.07529 [quant-ph]} \BibitemShut {NoStop}%
\bibitem [{\citenamefont {Liu}\ \emph {et~al.}(2021)\citenamefont {Liu}, \citenamefont {Li}, \citenamefont {Ragy}, \citenamefont {Zhao}, \citenamefont {Bai}, \citenamefont {Liu}, \citenamefont {Brown}, \citenamefont {Zhang}, \citenamefont {Colbeck}, \citenamefont {Fan}, \citenamefont {Zhang},\ and\ \citenamefont {Pan}}]{Liu2021}%
  \BibitemOpen
  \bibfield  {author} {\bibinfo {author} {\bibfnamefont {W.-Z.}\ \bibnamefont {Liu}}, \bibinfo {author} {\bibfnamefont {M.-H.}\ \bibnamefont {Li}}, \bibinfo {author} {\bibfnamefont {S.}~\bibnamefont {Ragy}}, \bibinfo {author} {\bibfnamefont {S.-R.}\ \bibnamefont {Zhao}}, \bibinfo {author} {\bibfnamefont {B.}~\bibnamefont {Bai}}, \bibinfo {author} {\bibfnamefont {Y.}~\bibnamefont {Liu}}, \bibinfo {author} {\bibfnamefont {P.~J.}\ \bibnamefont {Brown}}, \bibinfo {author} {\bibfnamefont {J.}~\bibnamefont {Zhang}}, \bibinfo {author} {\bibfnamefont {R.}~\bibnamefont {Colbeck}}, \bibinfo {author} {\bibfnamefont {J.}~\bibnamefont {Fan}}, \bibinfo {author} {\bibfnamefont {Q.}~\bibnamefont {Zhang}},\ and\ \bibinfo {author} {\bibfnamefont {J.-W.}\ \bibnamefont {Pan}},\ }\bibfield  {title} {\bibinfo {title} {Device-independent randomness expansion against quantum side information},\ }\href {https://doi.org/10.1038/s41567-020-01147-2} {\bibfield  {journal} {\bibinfo  {journal} {Nature Physics}\ }\textbf {\bibinfo {volume} {17}},\ \bibinfo {pages} {448} (\bibinfo {year} {2021})},\ \Eprint {https://arxiv.org/abs/1912.11159} {arXiv:1912.11159 [quant-ph]} \BibitemShut {NoStop}%
\bibitem [{\citenamefont {Bell}\ and\ \citenamefont {Aspect}(2004)}]{Bell2004}%
  \BibitemOpen
  \bibfield  {author} {\bibinfo {author} {\bibfnamefont {J.~S.}\ \bibnamefont {Bell}}\ and\ \bibinfo {author} {\bibfnamefont {A.}~\bibnamefont {Aspect}},\ }\href@noop {} {\emph {\bibinfo {title} {Speakable and Unspeakable in Quantum Mechanics: {C}ollected Papers on Quantum Philosophy}}},\ \bibinfo {edition} {2nd}\ ed.\ (\bibinfo  {publisher} {Cambridge University Press},\ \bibinfo {year} {2004})\BibitemShut {NoStop}%
\bibitem [{\citenamefont {Zukowski}\ \emph {et~al.}(1993)\citenamefont {Zukowski}, \citenamefont {Zeilinger}, \citenamefont {Horne},\ and\ \citenamefont {Ekert}}]{Zukowski1993}%
  \BibitemOpen
  \bibfield  {author} {\bibinfo {author} {\bibfnamefont {M.}~\bibnamefont {Zukowski}}, \bibinfo {author} {\bibfnamefont {A.}~\bibnamefont {Zeilinger}}, \bibinfo {author} {\bibfnamefont {M.~A.}\ \bibnamefont {Horne}},\ and\ \bibinfo {author} {\bibfnamefont {A.~K.}\ \bibnamefont {Ekert}},\ }\bibfield  {title} {\bibinfo {title} {“{E}vent-ready-detectors” {B}ell experiment via entanglement swapping},\ }\href {https://doi.org/10.1103/PhysRevLett.71.4287} {\bibfield  {journal} {\bibinfo  {journal} {Physical Review Letters}\ }\textbf {\bibinfo {volume} {71}},\ \bibinfo {pages} {4287} (\bibinfo {year} {1993})}\BibitemShut {NoStop}%
\bibitem [{\citenamefont {Sangouard}\ \emph {et~al.}(2011)\citenamefont {Sangouard}, \citenamefont {Sanguinetti}, \citenamefont {Curtz}, \citenamefont {Gisin}, \citenamefont {Thew},\ and\ \citenamefont {Zbinden}}]{Sangouard2011}%
  \BibitemOpen
  \bibfield  {author} {\bibinfo {author} {\bibfnamefont {N.}~\bibnamefont {Sangouard}}, \bibinfo {author} {\bibfnamefont {B.}~\bibnamefont {Sanguinetti}}, \bibinfo {author} {\bibfnamefont {N.}~\bibnamefont {Curtz}}, \bibinfo {author} {\bibfnamefont {N.}~\bibnamefont {Gisin}}, \bibinfo {author} {\bibfnamefont {R.}~\bibnamefont {Thew}},\ and\ \bibinfo {author} {\bibfnamefont {H.}~\bibnamefont {Zbinden}},\ }\bibfield  {title} {\bibinfo {title} {Faithful entanglement swapping based on sum-frequency generation},\ }\href {https://doi.org/10.1103/physrevlett.106.120403} {\bibfield  {journal} {\bibinfo  {journal} {Physical Review Letters}\ }\textbf {\bibinfo {volume} {106}},\ \bibinfo {pages} {120403} (\bibinfo {year} {2011})},\ \Eprint {https://arxiv.org/abs/1101.1009} {arXiv:1101.1009 [quant-ph]} \BibitemShut {NoStop}%
\bibitem [{\citenamefont {Curty}\ and\ \citenamefont {Moroder}(2011)}]{Curty2011}%
  \BibitemOpen
  \bibfield  {author} {\bibinfo {author} {\bibfnamefont {M.}~\bibnamefont {Curty}}\ and\ \bibinfo {author} {\bibfnamefont {T.}~\bibnamefont {Moroder}},\ }\bibfield  {title} {\bibinfo {title} {Heralded-qubit amplifiers for practical device-independent quantum key distribution},\ }\href {https://doi.org/10.1103/physreva.84.010304} {\bibfield  {journal} {\bibinfo  {journal} {Physical Review A}\ }\textbf {\bibinfo {volume} {84}},\ \bibinfo {pages} {010304} (\bibinfo {year} {2011})},\ \Eprint {https://arxiv.org/abs/1105.2573} {arXiv:1105.2573 [quant-ph]} \BibitemShut {NoStop}%
\bibitem [{\citenamefont {Gisin}\ \emph {et~al.}(2010)\citenamefont {Gisin}, \citenamefont {Pironio},\ and\ \citenamefont {Sangouard}}]{Gisin2010}%
  \BibitemOpen
  \bibfield  {author} {\bibinfo {author} {\bibfnamefont {N.}~\bibnamefont {Gisin}}, \bibinfo {author} {\bibfnamefont {S.}~\bibnamefont {Pironio}},\ and\ \bibinfo {author} {\bibfnamefont {N.}~\bibnamefont {Sangouard}},\ }\bibfield  {title} {\bibinfo {title} {Proposal for implementing device-independent quantum key distribution based on a heralded qubit amplifier},\ }\href {https://doi.org/https://doi.org/10.1103/PhysRevLett.105.070501} {\bibfield  {journal} {\bibinfo  {journal} {Physical Review Letters}\ }\textbf {\bibinfo {volume} {105}},\ \bibinfo {pages} {070501} (\bibinfo {year} {2010})},\ \Eprint {https://arxiv.org/abs/1003.0635} {arXiv:1003.0635 [quant-ph]} \BibitemShut {NoStop}%
\bibitem [{\citenamefont {Cabello}\ and\ \citenamefont {Sciarrino}(2012)}]{Cabello2012}%
  \BibitemOpen
  \bibfield  {author} {\bibinfo {author} {\bibfnamefont {A.}~\bibnamefont {Cabello}}\ and\ \bibinfo {author} {\bibfnamefont {F.}~\bibnamefont {Sciarrino}},\ }\bibfield  {title} {\bibinfo {title} {Loophole-free {B}ell test based on local precertification of photon’s presence},\ }\href {https://doi.org/10.1103/PhysRevX.2.021010} {\bibfield  {journal} {\bibinfo  {journal} {Physical Review X}\ }\textbf {\bibinfo {volume} {2}},\ \bibinfo {pages} {021010} (\bibinfo {year} {2012})}\BibitemShut {NoStop}%
\bibitem [{\citenamefont {Azuma}\ \emph {et~al.}(2023)\citenamefont {Azuma}, \citenamefont {Economou}, \citenamefont {Elkouss}, \citenamefont {Hilaire}, \citenamefont {Jiang}, \citenamefont {Lo},\ and\ \citenamefont {Tzitrin}}]{Azuma2023}%
  \BibitemOpen
  \bibfield  {author} {\bibinfo {author} {\bibfnamefont {K.}~\bibnamefont {Azuma}}, \bibinfo {author} {\bibfnamefont {S.~E.}\ \bibnamefont {Economou}}, \bibinfo {author} {\bibfnamefont {D.}~\bibnamefont {Elkouss}}, \bibinfo {author} {\bibfnamefont {P.}~\bibnamefont {Hilaire}}, \bibinfo {author} {\bibfnamefont {L.}~\bibnamefont {Jiang}}, \bibinfo {author} {\bibfnamefont {H.-K.}\ \bibnamefont {Lo}},\ and\ \bibinfo {author} {\bibfnamefont {I.}~\bibnamefont {Tzitrin}},\ }\bibfield  {title} {\bibinfo {title} {Quantum repeaters: {F}rom quantum networks to the quantum internet},\ }\href {https://doi.org/10.1103/revmodphys.95.045006} {\bibfield  {journal} {\bibinfo  {journal} {Reviews of Modern Physics}\ }\textbf {\bibinfo {volume} {95}},\ \bibinfo {pages} {045006} (\bibinfo {year} {2023})},\ \Eprint {https://arxiv.org/abs/2212.10820} {arXiv:2212.10820 [quant-ph]} \BibitemShut {NoStop}%
\bibitem [{\citenamefont {Chaturvedi}\ \emph {et~al.}(2024)\citenamefont {Chaturvedi}, \citenamefont {Viola},\ and\ \citenamefont {Pawłowski}}]{Chaturvedi2022}%
  \BibitemOpen
  \bibfield  {author} {\bibinfo {author} {\bibfnamefont {A.}~\bibnamefont {Chaturvedi}}, \bibinfo {author} {\bibfnamefont {G.}~\bibnamefont {Viola}},\ and\ \bibinfo {author} {\bibfnamefont {M.}~\bibnamefont {Pawłowski}},\ }\bibfield  {title} {\bibinfo {title} {Extending loophole-free nonlocal correlations to arbitrarily large distances},\ }\href {https://doi.org/10.1038/s41534-023-00799-1} {\bibfield  {journal} {\bibinfo  {journal} {npj Quantum Information}\ }\textbf {\bibinfo {volume} {10}},\ \bibinfo {pages} {7} (\bibinfo {year} {2024})},\ \Eprint {https://arxiv.org/abs/2211.14231} {arXiv:2211.14231 [quant-ph]} \BibitemShut {NoStop}%
\bibitem [{\citenamefont {Lobo}\ \emph {et~al.}(2024)\citenamefont {Lobo}, \citenamefont {Pauwels},\ and\ \citenamefont {Pironio}}]{Lobo2023}%
  \BibitemOpen
  \bibfield  {author} {\bibinfo {author} {\bibfnamefont {E.~P.}\ \bibnamefont {Lobo}}, \bibinfo {author} {\bibfnamefont {J.}~\bibnamefont {Pauwels}},\ and\ \bibinfo {author} {\bibfnamefont {S.}~\bibnamefont {Pironio}},\ }\bibfield  {title} {\bibinfo {title} {Certifying long-range quantum correlations through routed {B}ell tests},\ }\href {https://doi.org/10.22331/q-2024-05-02-1332} {\bibfield  {journal} {\bibinfo  {journal} {Quantum}\ }\textbf {\bibinfo {volume} {8}},\ \bibinfo {pages} {1332} (\bibinfo {year} {2024})},\ \Eprint {https://arxiv.org/abs/2310.07484} {arXiv:2310.07484 [quant-ph]} \BibitemShut {NoStop}%
\bibitem [{\citenamefont {Massar}\ and\ \citenamefont {Pironio}(2003)}]{Massar2003}%
  \BibitemOpen
  \bibfield  {author} {\bibinfo {author} {\bibfnamefont {S.}~\bibnamefont {Massar}}\ and\ \bibinfo {author} {\bibfnamefont {S.}~\bibnamefont {Pironio}},\ }\bibfield  {title} {\bibinfo {title} {Violation of local realism versus detection efficiency},\ }\href {https://doi.org/10.1103/physreva.68.062109} {\bibfield  {journal} {\bibinfo  {journal} {Physical Review A}\ }\textbf {\bibinfo {volume} {68}},\ \bibinfo {pages} {062109} (\bibinfo {year} {2003})},\ \Eprint {https://arxiv.org/abs/quant-ph/0210103} {arXiv:quant-ph/0210103} \BibitemShut {NoStop}%
\bibitem [{\citenamefont {Navascu\'es}\ \emph {et~al.}(2007)\citenamefont {Navascu\'es}, \citenamefont {Pironio},\ and\ \citenamefont {Ac\'{\i}n}}]{Navascues2007}%
  \BibitemOpen
  \bibfield  {author} {\bibinfo {author} {\bibfnamefont {M.}~\bibnamefont {Navascu\'es}}, \bibinfo {author} {\bibfnamefont {S.}~\bibnamefont {Pironio}},\ and\ \bibinfo {author} {\bibfnamefont {A.}~\bibnamefont {Ac\'{\i}n}},\ }\bibfield  {title} {\bibinfo {title} {Bounding the set of quantum correlations},\ }\href {https://doi.org/10.1103/physrevlett.98.010401} {\bibfield  {journal} {\bibinfo  {journal} {Physical Review Letters}\ }\textbf {\bibinfo {volume} {98}},\ \bibinfo {pages} {010401} (\bibinfo {year} {2007})},\ \Eprint {https://arxiv.org/abs/quant-ph/0607119} {quant-ph/0607119} \BibitemShut {NoStop}%
\bibitem [{\citenamefont {Navascu{\'e}s}\ \emph {et~al.}(2008)\citenamefont {Navascu{\'e}s}, \citenamefont {Pironio},\ and\ \citenamefont {Ac{\'\i}n}}]{Navascues2008}%
  \BibitemOpen
  \bibfield  {author} {\bibinfo {author} {\bibfnamefont {M.}~\bibnamefont {Navascu{\'e}s}}, \bibinfo {author} {\bibfnamefont {S.}~\bibnamefont {Pironio}},\ and\ \bibinfo {author} {\bibfnamefont {A.}~\bibnamefont {Ac{\'\i}n}},\ }\bibfield  {title} {\bibinfo {title} {A convergent hierarchy of semidefinite programs characterizing the set of quantum correlations},\ }\href {https://doi.org/10.1088/1367-2630/10/7/073013} {\bibfield  {journal} {\bibinfo  {journal} {New Journal of Physics}\ }\textbf {\bibinfo {volume} {10}},\ \bibinfo {pages} {073013} (\bibinfo {year} {2008})},\ \Eprint {https://arxiv.org/abs/0803.4290} {arXiv:0803.4290 [quant-ph]} \BibitemShut {NoStop}%
\bibitem [{\citenamefont {Pironio}\ \emph {et~al.}(2010{\natexlab{a}})\citenamefont {Pironio}, \citenamefont {Navascu{\'e}s},\ and\ \citenamefont {Acin}}]{Pironio2010a}%
  \BibitemOpen
  \bibfield  {author} {\bibinfo {author} {\bibfnamefont {S.}~\bibnamefont {Pironio}}, \bibinfo {author} {\bibfnamefont {M.}~\bibnamefont {Navascu{\'e}s}},\ and\ \bibinfo {author} {\bibfnamefont {A.}~\bibnamefont {Acin}},\ }\bibfield  {title} {\bibinfo {title} {Convergent relaxations of polynomial optimization problems with noncommuting variables},\ }\href {https://doi.org/10.1137/090760155} {\bibfield  {journal} {\bibinfo  {journal} {SIOPT}\ }\textbf {\bibinfo {volume} {20}},\ \bibinfo {pages} {2157} (\bibinfo {year} {2010}{\natexlab{a}})},\ \Eprint {https://arxiv.org/abs/0903.4368} {arXiv:0903.4368 [math.OC]} \BibitemShut {NoStop}%
\bibitem [{\citenamefont {Brown}\ \emph {et~al.}(2021)\citenamefont {Brown}, \citenamefont {Fawzi},\ and\ \citenamefont {Fawzi}}]{Brown2021}%
  \BibitemOpen
  \bibfield  {author} {\bibinfo {author} {\bibfnamefont {P.}~\bibnamefont {Brown}}, \bibinfo {author} {\bibfnamefont {H.}~\bibnamefont {Fawzi}},\ and\ \bibinfo {author} {\bibfnamefont {O.}~\bibnamefont {Fawzi}},\ }\bibfield  {title} {\bibinfo {title} {Computing conditional entropies for quantum correlations},\ }\href {https://doi.org/10.1038/s41467-020-20018-1} {\bibfield  {journal} {\bibinfo  {journal} {Nature communications}\ }\textbf {\bibinfo {volume} {12}},\ \bibinfo {pages} {575} (\bibinfo {year} {2021})},\ \Eprint {https://arxiv.org/abs/2007.12575} {arXiv:2007.12575 [quant-ph]} \BibitemShut {NoStop}%
\bibitem [{\citenamefont {Pironio}\ \emph {et~al.}(2010{\natexlab{b}})\citenamefont {Pironio}, \citenamefont {Ac{\'\i}n}, \citenamefont {Massar}, \citenamefont {de~La~Giroday}, \citenamefont {Matsukevich}, \citenamefont {Maunz}, \citenamefont {Olmschenk}, \citenamefont {Hayes}, \citenamefont {Luo}, \citenamefont {Manning} \emph {et~al.}}]{Pironio2010}%
  \BibitemOpen
  \bibfield  {author} {\bibinfo {author} {\bibfnamefont {S.}~\bibnamefont {Pironio}}, \bibinfo {author} {\bibfnamefont {A.}~\bibnamefont {Ac{\'\i}n}}, \bibinfo {author} {\bibfnamefont {S.}~\bibnamefont {Massar}}, \bibinfo {author} {\bibfnamefont {A.~B.}\ \bibnamefont {de~La~Giroday}}, \bibinfo {author} {\bibfnamefont {D.~N.}\ \bibnamefont {Matsukevich}}, \bibinfo {author} {\bibfnamefont {P.}~\bibnamefont {Maunz}}, \bibinfo {author} {\bibfnamefont {S.}~\bibnamefont {Olmschenk}}, \bibinfo {author} {\bibfnamefont {D.}~\bibnamefont {Hayes}}, \bibinfo {author} {\bibfnamefont {L.}~\bibnamefont {Luo}}, \bibinfo {author} {\bibfnamefont {T.~A.}\ \bibnamefont {Manning}}, \emph {et~al.},\ }\bibfield  {title} {\bibinfo {title} {Random numbers certified by {B}ell's theorem},\ }\href {https://doi.org/10.1038/nature09008} {\bibfield  {journal} {\bibinfo  {journal} {Nature}\ }\textbf {\bibinfo {volume} {464}},\ \bibinfo {pages} {1021} (\bibinfo {year} {2010}{\natexlab{b}})},\ \Eprint {https://arxiv.org/abs/0911.3427} {arXiv:0911.3427 [quant-ph]} \BibitemShut {NoStop}%
\bibitem [{\citenamefont {Maurer}\ and\ \citenamefont {Wolf}(1999)}]{maurer1999unconditionally}%
  \BibitemOpen
  \bibfield  {author} {\bibinfo {author} {\bibfnamefont {U.~M.}\ \bibnamefont {Maurer}}\ and\ \bibinfo {author} {\bibfnamefont {S.}~\bibnamefont {Wolf}},\ }\bibfield  {title} {\bibinfo {title} {Unconditionally secure key agreement and the intrinsic conditional information},\ }\href {https://doi.org/10.1109/18.748999} {\bibfield  {journal} {\bibinfo  {journal} {IEEE Transactions on Information Theory}\ }\textbf {\bibinfo {volume} {45}},\ \bibinfo {pages} {499} (\bibinfo {year} {1999})}\BibitemShut {NoStop}%
\bibitem [{\citenamefont {Dupuis}\ \emph {et~al.}(2020)\citenamefont {Dupuis}, \citenamefont {Fawzi},\ and\ \citenamefont {Renner}}]{dupuis2020entropy}%
  \BibitemOpen
  \bibfield  {author} {\bibinfo {author} {\bibfnamefont {F.}~\bibnamefont {Dupuis}}, \bibinfo {author} {\bibfnamefont {O.}~\bibnamefont {Fawzi}},\ and\ \bibinfo {author} {\bibfnamefont {R.}~\bibnamefont {Renner}},\ }\bibfield  {title} {\bibinfo {title} {Entropy accumulation},\ }\href {https://doi.org/10.1007/s00220-020-03839-5} {\bibfield  {journal} {\bibinfo  {journal} {Communications in Mathematical Physics}\ }\textbf {\bibinfo {volume} {379}},\ \bibinfo {pages} {867} (\bibinfo {year} {2020})},\ \Eprint {https://arxiv.org/abs/1607.01796} {arXiv:1607.01796 [quant-ph]} \BibitemShut {NoStop}%
\bibitem [{\citenamefont {Arnon-Friedman}\ \emph {et~al.}(2018)\citenamefont {Arnon-Friedman}, \citenamefont {Dupuis}, \citenamefont {Fawzi}, \citenamefont {Renner},\ and\ \citenamefont {Vidick}}]{arnon2018practical}%
  \BibitemOpen
  \bibfield  {author} {\bibinfo {author} {\bibfnamefont {R.}~\bibnamefont {Arnon-Friedman}}, \bibinfo {author} {\bibfnamefont {F.}~\bibnamefont {Dupuis}}, \bibinfo {author} {\bibfnamefont {O.}~\bibnamefont {Fawzi}}, \bibinfo {author} {\bibfnamefont {R.}~\bibnamefont {Renner}},\ and\ \bibinfo {author} {\bibfnamefont {T.}~\bibnamefont {Vidick}},\ }\bibfield  {title} {\bibinfo {title} {Practical device-independent quantum cryptography via entropy accumulation},\ }\href {https://doi.org/10.1038/s41467-017-02307-4} {\bibfield  {journal} {\bibinfo  {journal} {Nature communications}\ }\textbf {\bibinfo {volume} {9}},\ \bibinfo {pages} {459} (\bibinfo {year} {2018})}\BibitemShut {NoStop}%
\bibitem [{\citenamefont {Arnon-Friedman}\ \emph {et~al.}(2019)\citenamefont {Arnon-Friedman}, \citenamefont {Renner},\ and\ \citenamefont {Vidick}}]{arnon2019simple}%
  \BibitemOpen
  \bibfield  {author} {\bibinfo {author} {\bibfnamefont {R.}~\bibnamefont {Arnon-Friedman}}, \bibinfo {author} {\bibfnamefont {R.}~\bibnamefont {Renner}},\ and\ \bibinfo {author} {\bibfnamefont {T.}~\bibnamefont {Vidick}},\ }\bibfield  {title} {\bibinfo {title} {Simple and tight device-independent security proofs},\ }\href {https://doi.org/10.1137/18m1174726} {\bibfield  {journal} {\bibinfo  {journal} {SIAM Journal on Computing}\ }\textbf {\bibinfo {volume} {48}},\ \bibinfo {pages} {181} (\bibinfo {year} {2019})},\ \Eprint {https://arxiv.org/abs/1607.01797} {arXiv:1607.01797 [quant-ph]} \BibitemShut {NoStop}%
\bibitem [{\citenamefont {Metger}\ \emph {et~al.}(2022)\citenamefont {Metger}, \citenamefont {Fawzi}, \citenamefont {Sutter},\ and\ \citenamefont {Renner}}]{Metger2022}%
  \BibitemOpen
  \bibfield  {author} {\bibinfo {author} {\bibfnamefont {T.}~\bibnamefont {Metger}}, \bibinfo {author} {\bibfnamefont {O.}~\bibnamefont {Fawzi}}, \bibinfo {author} {\bibfnamefont {D.}~\bibnamefont {Sutter}},\ and\ \bibinfo {author} {\bibfnamefont {R.}~\bibnamefont {Renner}},\ }\bibfield  {title} {\bibinfo {title} {Generalised entropy accumulation},\ }in\ \href {https://doi.org/10.1109/focs54457.2022.00085} {\emph {\bibinfo {booktitle} {2022 IEEE 63rd Annual Symposium on Foundations of Computer Science (FOCS)}}}\ (\bibinfo {year} {2022})\ pp.\ \bibinfo {pages} {844--850},\ \Eprint {https://arxiv.org/abs/2203.04989} {arXiv:2203.04989 [quant-ph]} \BibitemShut {NoStop}%
\bibitem [{\citenamefont {Devetak}\ and\ \citenamefont {Winter}(2005)}]{Devetak2005}%
  \BibitemOpen
  \bibfield  {author} {\bibinfo {author} {\bibfnamefont {I.}~\bibnamefont {Devetak}}\ and\ \bibinfo {author} {\bibfnamefont {A.}~\bibnamefont {Winter}},\ }\bibfield  {title} {\bibinfo {title} {Distillation of secret key and entanglement from quantum states},\ }\href {https://doi.org/https://doi.org/10.1098/rspa.2004.1372} {\bibfield  {journal} {\bibinfo  {journal} {Proceedings of the Royal Society A: Mathematical, Physical and Engineering Sciences}\ }\textbf {\bibinfo {volume} {461}},\ \bibinfo {pages} {207} (\bibinfo {year} {2005})},\ \Eprint {https://arxiv.org/abs/quant-ph/0306078} {arXiv:quant-ph/0306078} \BibitemShut {NoStop}%
\bibitem [{\citenamefont {Masini}\ and\ \citenamefont {Sarkar}(2024)}]{masini2024one}%
  \BibitemOpen
  \bibfield  {author} {\bibinfo {author} {\bibfnamefont {M.}~\bibnamefont {Masini}}\ and\ \bibinfo {author} {\bibfnamefont {S.}~\bibnamefont {Sarkar}},\ }\href@noop {} {\bibinfo {title} {One-sided di-qkd secure against coherent attacks over long distances}} (\bibinfo {year} {2024}),\ \Eprint {https://arxiv.org/abs/2403.11850} {arXiv:2403.11850 [quant-ph]} \BibitemShut {NoStop}%
\bibitem [{\citenamefont {Mayers}(2001)}]{mayers2001unconditional}%
  \BibitemOpen
  \bibfield  {author} {\bibinfo {author} {\bibfnamefont {D.}~\bibnamefont {Mayers}},\ }\bibfield  {title} {\bibinfo {title} {Unconditional security in quantum cryptography},\ }\href {https://doi.org/https://doi.org/10.1145/382780.382781} {\bibfield  {journal} {\bibinfo  {journal} {Journal of the ACM}\ }\textbf {\bibinfo {volume} {48}},\ \bibinfo {pages} {351} (\bibinfo {year} {2001})},\ \Eprint {https://arxiv.org/abs/quant-ph/9802025} {arXiv:quant-ph/9802025 [quant-ph]} \BibitemShut {NoStop}%
\bibitem [{\citenamefont {Berta}\ \emph {et~al.}(2010)\citenamefont {Berta}, \citenamefont {Christandl}, \citenamefont {Colbeck}, \citenamefont {Renes},\ and\ \citenamefont {Renner}}]{berta2010uncertainty}%
  \BibitemOpen
  \bibfield  {author} {\bibinfo {author} {\bibfnamefont {M.}~\bibnamefont {Berta}}, \bibinfo {author} {\bibfnamefont {M.}~\bibnamefont {Christandl}}, \bibinfo {author} {\bibfnamefont {R.}~\bibnamefont {Colbeck}}, \bibinfo {author} {\bibfnamefont {J.~M.}\ \bibnamefont {Renes}},\ and\ \bibinfo {author} {\bibfnamefont {R.}~\bibnamefont {Renner}},\ }\bibfield  {title} {\bibinfo {title} {The uncertainty principle in the presence of quantum memory},\ }\href {https://doi.org/10.1038/nphys1734} {\bibfield  {journal} {\bibinfo  {journal} {Nature Physics}\ }\textbf {\bibinfo {volume} {6}},\ \bibinfo {pages} {659} (\bibinfo {year} {2010})},\ \Eprint {https://arxiv.org/abs/0909.0950} {arXiv:0909.0950 [quant-ph]} \BibitemShut {NoStop}%
\bibitem [{\citenamefont {Woodhead}(2016)}]{woodhead2016semi}%
  \BibitemOpen
  \bibfield  {author} {\bibinfo {author} {\bibfnamefont {E.}~\bibnamefont {Woodhead}},\ }\bibfield  {title} {\bibinfo {title} {Semi device independence of the {BB84} protocol},\ }\href {https://doi.org/10.1088/1367-2630/18/5/055010} {\bibfield  {journal} {\bibinfo  {journal} {New Journal of Physics}\ }\textbf {\bibinfo {volume} {18}},\ \bibinfo {pages} {055010} (\bibinfo {year} {2016})},\ \Eprint {https://arxiv.org/abs/1512.03387} {arXiv:1512.03387 [quant-ph]} \BibitemShut {NoStop}%
\bibitem [{\citenamefont {Woodhead}\ and\ \citenamefont {Pironio}(2015)}]{woodhead2015secrecy}%
  \BibitemOpen
  \bibfield  {author} {\bibinfo {author} {\bibfnamefont {E.}~\bibnamefont {Woodhead}}\ and\ \bibinfo {author} {\bibfnamefont {S.}~\bibnamefont {Pironio}},\ }\bibfield  {title} {\bibinfo {title} {Secrecy in prepare-and-measure {Clauser-Horne-Shimony-Holt} tests with a qubit bound},\ }\href {https://doi.org/10.1103/PhysRevLett.115.150501} {\bibfield  {journal} {\bibinfo  {journal} {Physical Review Letters}\ }\textbf {\bibinfo {volume} {115}},\ \bibinfo {pages} {150501} (\bibinfo {year} {2015})},\ \Eprint {https://arxiv.org/abs/1507.02889} {arXiv:1507.02889 [quant-ph]} \BibitemShut {NoStop}%
\bibitem [{\citenamefont {Woodhead}\ \emph {et~al.}(2012)\citenamefont {Woodhead}, \citenamefont {Lim},\ and\ \citenamefont {Pironio}}]{woodhead2012semi}%
  \BibitemOpen
  \bibfield  {author} {\bibinfo {author} {\bibfnamefont {E.}~\bibnamefont {Woodhead}}, \bibinfo {author} {\bibfnamefont {C.~C.~W.}\ \bibnamefont {Lim}},\ and\ \bibinfo {author} {\bibfnamefont {S.}~\bibnamefont {Pironio}},\ }\bibfield  {title} {\bibinfo {title} {Semi-device-independent {QKD} based on {BB84} and a {CHSH}-type estimation},\ }in\ \href {https://doi.org/10.1007/978-3-642-35656-8_9} {\emph {\bibinfo {booktitle} {Conference on Quantum Computation, Communication, and Cryptography}}}\ (\bibinfo {organization} {Springer},\ \bibinfo {year} {2012})\ pp.\ \bibinfo {pages} {107--115}\BibitemShut {NoStop}%
\bibitem [{\citenamefont {Le~Roy-Deloison}\ and\ \citenamefont {Lobo}()}]{Git}%
  \BibitemOpen
  \bibfield  {author} {\bibinfo {author} {\bibfnamefont {T.}~\bibnamefont {Le~Roy-Deloison}}\ and\ \bibinfo {author} {\bibfnamefont {E.~P.}\ \bibnamefont {Lobo}},\ }\href@noop {} {\bibinfo {title} {Plots and scripts to reproduce routed {DIQKD} results}},\ \bibinfo {howpublished} {\url{https://github.com/tristan-le-roy/Routed-DIQKD}}\BibitemShut {NoStop}%
\bibitem [{\citenamefont {Shor}\ and\ \citenamefont {Preskill}(2000)}]{Shor2000}%
  \BibitemOpen
  \bibfield  {author} {\bibinfo {author} {\bibfnamefont {P.~W.}\ \bibnamefont {Shor}}\ and\ \bibinfo {author} {\bibfnamefont {J.}~\bibnamefont {Preskill}},\ }\bibfield  {title} {\bibinfo {title} {Simple proof of security of the {BB84} quantum key distribution protocol},\ }\href {https://doi.org/10.1103/physrevlett.85.441} {\bibfield  {journal} {\bibinfo  {journal} {Physical Review Letters}\ }\textbf {\bibinfo {volume} {85}},\ \bibinfo {pages} {441} (\bibinfo {year} {2000})},\ \Eprint {https://arxiv.org/abs/quant-ph/0003004} {arXiv:quant-ph/0003004} \BibitemShut {NoStop}%
\bibitem [{\citenamefont {Masini}\ \emph {et~al.}(2024)\citenamefont {Masini}, \citenamefont {Ioannou}, \citenamefont {Brunner}, \citenamefont {Pironio},\ and\ \citenamefont {Sekatski}}]{masini2024joint}%
  \BibitemOpen
  \bibfield  {author} {\bibinfo {author} {\bibfnamefont {M.}~\bibnamefont {Masini}}, \bibinfo {author} {\bibfnamefont {M.}~\bibnamefont {Ioannou}}, \bibinfo {author} {\bibfnamefont {N.}~\bibnamefont {Brunner}}, \bibinfo {author} {\bibfnamefont {S.}~\bibnamefont {Pironio}},\ and\ \bibinfo {author} {\bibfnamefont {P.}~\bibnamefont {Sekatski}},\ }\bibfield  {title} {\bibinfo {title} {Joint-measurability and quantum communication with untrusted devices},\ }\href@noop {} {\bibfield  {journal} {\bibinfo  {journal} {Quantum}\ }\textbf {\bibinfo {volume} {8}},\ \bibinfo {pages} {1574} (\bibinfo {year} {2024})},\ \Eprint {https://arxiv.org/abs/2403.14785} {arXiv:2403.14785 [quant-ph]} \BibitemShut {NoStop}%
\bibitem [{\citenamefont {Lim}\ \emph {et~al.}(2013)\citenamefont {Lim}, \citenamefont {Portmann}, \citenamefont {Tomamichel}, \citenamefont {Renner},\ and\ \citenamefont {Gisin}}]{lim2013device}%
  \BibitemOpen
  \bibfield  {author} {\bibinfo {author} {\bibfnamefont {C.~C.~W.}\ \bibnamefont {Lim}}, \bibinfo {author} {\bibfnamefont {C.}~\bibnamefont {Portmann}}, \bibinfo {author} {\bibfnamefont {M.}~\bibnamefont {Tomamichel}}, \bibinfo {author} {\bibfnamefont {R.}~\bibnamefont {Renner}},\ and\ \bibinfo {author} {\bibfnamefont {N.}~\bibnamefont {Gisin}},\ }\bibfield  {title} {\bibinfo {title} {Device-independent quantum key distribution with local bell test},\ }\href {https://doi.org/10.1103/PhysRevX.3.031006} {\bibfield  {journal} {\bibinfo  {journal} {Phys. Rev. X}\ }\textbf {\bibinfo {volume} {3}},\ \bibinfo {pages} {031006} (\bibinfo {year} {2013})},\ \Eprint {https://arxiv.org/abs/1208.0023} {arXiv:1208.0023 [quant-ph]} \BibitemShut {NoStop}%
\bibitem [{\citenamefont {Tan}\ and\ \citenamefont {Wolf}(2024)}]{Tan2024}%
  \BibitemOpen
  \bibfield  {author} {\bibinfo {author} {\bibfnamefont {E.~Y.-Z.}\ \bibnamefont {Tan}}\ and\ \bibinfo {author} {\bibfnamefont {R.}~\bibnamefont {Wolf}},\ }\bibfield  {title} {\bibinfo {title} {Entropy bounds for device-independent quantum key distribution with local {B}ell test},\ }\href {https://doi.org/10.1103/PhysRevLett.133.120803} {\bibfield  {journal} {\bibinfo  {journal} {Physical Review Letters}\ }\textbf {\bibinfo {volume} {133}},\ \bibinfo {pages} {120803} (\bibinfo {year} {2024})},\ \Eprint {https://arxiv.org/abs/2404.00792} {arXiv:2404.00792 [quant-ph]} \BibitemShut {NoStop}%
\bibitem [{\citenamefont {Brown}\ \emph {et~al.}(2024)\citenamefont {Brown}, \citenamefont {Fawzi},\ and\ \citenamefont {Fawzi}}]{Brown2021a}%
  \BibitemOpen
  \bibfield  {author} {\bibinfo {author} {\bibfnamefont {P.}~\bibnamefont {Brown}}, \bibinfo {author} {\bibfnamefont {H.}~\bibnamefont {Fawzi}},\ and\ \bibinfo {author} {\bibfnamefont {O.}~\bibnamefont {Fawzi}},\ }\bibfield  {title} {\bibinfo {title} {Device-independent lower bounds on the conditional von {N}eumann entropy},\ }\href {https://doi.org/https://doi.org/10.22331/q-2024-08-27-1445} {\bibfield  {journal} {\bibinfo  {journal} {Quantum}\ }\textbf {\bibinfo {volume} {8}},\ \bibinfo {pages} {1445} (\bibinfo {year} {2024})},\ \Eprint {https://arxiv.org/abs/2106.13692} {arXiv:2106.13692 [quant-ph]} \BibitemShut {NoStop}%
\end{thebibliography}%

\end{document}